\documentclass[12pt]{JHEP3} 
\input colordvi

\title{\textRed{Before and After: How has the SNO neutral current measurement
changed things?}\textBlack}

\author{John N. Bahcall\\
  School of Natural Sciences, Institute for Advanced Study, Princeton,
  NJ 08540\\
    E-mail: \email{jnb@ias.edu}}
\author{M. C. Gonzalez-Garcia \\
  Theory Division, CERN, CH-1211, Geneva 23, Switzerland,\\
  Y.I.T.P., SUNY at Stony Brook, Stony Brook,NY 11794-3840\\
and IFIC, Universitat de Val\`encia -- C.S.I.C., Apt 22085, 46071
  Val\`encia, Spain\\
    E-mail: \email{concepcion.gonzalez-garcia@cern.ch}}
\author{Carlos Pe\~na-Garay\\
        IFIC, Universitat de Val\`encia -- C.S.I.C., Apt 22085, 46071
  Val\`encia, Spain and\\
  Theory Division, CERN, CH-1211, Geneva 23, Switzerland\\
       E-mail: \email{penya@ific.uv.es}}

\preprint{CERN-TH/2002-094\\
hep-ph/0204314}

\abstract{\textBlue  We present ``Before and After" global
oscillation solutions, as well as predicted ``Before and After"
values and ranges for ten future solar neutrino observables (for
BOREXINO, KamLAND, SNO, and a generic $p-p$ neutrino detector).
The ``Before'' case includes all solar neutrino data (and some
theoretical improvements) available prior to April 20, 2002 and
the ``After'' case includes, in addition, the new SNO data on the
CC, NC, and day-night asymmetry. We have performed global analyses
using the full SNO day-night energy spectrum and, alternatively,
using just the SNO NC and CC rates and the day-night asymmetry.
The LMA solution is the only currently allowed MSW oscillation
solution at $\sim 99$\% CL. The LOW solution is allowed only at
more than $2.5\sigma$, SMA is now excluded at $3.7\sigma$ or
$4.7\sigma$ depending upon analysis strategy, and pure sterile
oscillations are excluded at more than $4.7\sigma$. Small mixing
angles are ``out'' (pure sterile is ``way out''); MSW with large
mixing angles is definitely ``in.'' Vacuum oscillations are
allowed at $3\sigma$, but not at $2\sigma$. Precise maximal mixing
is excluded at $3.2\sigma$ for MSW solutions and at  more than
$2.8\sigma$ for vacuum solutions. Most of the predicted values for
future observables for the BOREXINO, KamLAND, and future SNO
measurements are changed only by minor amounts by the inclusion of
the recent SNO data. In order to test the robustness of the
allowed neutrino oscillation regions that are inferred from the
measurements and the predicted values for future solar neutrino
observables, we have carried out calculations using a variety of
strategies for analyzing the SNO and other experimental data. }


\begin{document}
\input psfig
\textBlack

\section{Introduction}
\label{sec:introduction}

The goal of this paper is to assess the impact of recent SNO
measurements ~\cite{snonc,snodaynight,snourl} on the allowed
regions of neutrino oscillation parameters and on the predicted
values of the most important future solar neutrino observables.
The SNO collaboration has reported a neutral current (NC)
measurement of the active $^8$B solar neutrino flux and related
measurements of the day-night asymmetry, as well as improved
determinations of the charged current (CC) and neutrino-electron
scattering rate.

We are concerned that global solutions for oscillation parameters
depend upon the assumption that the errors are well understood for
all of the reported measurements in the chlorine, gallium,
Super-Kamiokande, and SNO experiments. There are many cases in the
past for which similar assumptions have proved misleading.
Therefore, we focus our study on determining the robustness of our
conclusions regarding the currently allowed oscillation parameters
and the predicted values of new solar neutrino observables. We
test the robustness of the conclusions about allowed oscillation
parameters by using three different analysis strategies. We also
compute predicted values for future solar neutrino observables
with the oscillation parameters that are currently allowed as well
with the oscillation parameters that were allowed before the
recent SNO measurements. In addition, we treat the SNO data in two
different ways: a) with the aid of a two-step process (discussed
in section~\ref{subsec:aficionados}) and b) using the full SNO
day-night energy spectrum (see section~\ref{subsec:aficionadosFF})
used by the SNO collaboration. In our view, a necessary condition
for a result to be regarded as "robust" is that the result not
change significantly as we vary the analysis procedures among the
different plausible possibilities listed above.

Perhaps the most remarkable and encouraging result of our analysis
is that the allowed regions for the oscillation parameters,
$\Delta m^2$ and $\tan ^2 \theta$, the predicted values of ten
future solar neutrino observables, and the inferred total $^8$B
neutrino flux are all rather robust with respect to the choices
among the different analysis procedures. The inferences from the
available data are relatively independent of the details of the
analysis procedures. This conclusion will be justified
quantitatively in section~\ref{sec:fullsno} and summarized in
section~\ref{sec:discuss}.

We begin in section~\ref{sec:global} by deriving, using the
two-step procedure for the SNO data, the currently allowed regions
in neutrino oscillation space that are obtained with three
different analysis strategies (see figure~\ref{fig:global3}), each
strategy previously advocated by a different set of authors. The
global solutions obtained here are calculated using the methods
described in our recent paper~\cite{robust} (see especially
section~3.3 of ref.~\cite{robust}) but also include some
refinements in addition to the new SNO data. For example, we take
account of the energy dependence and correlations of the errors in
the neutrino absorption cross sections for the chlorine and
gallium solar neutrino experiments as described in the Appendix of
ref.~\cite{sterile}.  We also include the recently reported
SAGE~\cite{sage2002} data for 11 years of observation and the
zenith angle-recoil energy spectrum data presented by the
Super-Kamiokande collaboration after $1496$ days of
observations~\cite{smy2002} . Where required, we use the predicted
fluxes and their errors from the BP00 standard solar
model~\cite{bp2000}.

We present in section~\ref{sec:beforeafter}, and especially in
figure \ref{fig:beforeafter},\, a ``Before and After'' comparison
of the globally allowed neutrino oscillation solutions (see
figure~\ref{fig:beforeafter}). In the ``Before'' case, we use all
the solar neutrino data (see
refs.~\cite{sno2001,chlorine,gallex,gno,sage,superk,chooz}) that
were published or had appeared publicly before April 20, 2002, the
date that the SNO NC and day-night asymmetry were first published.
In the ``After'' case, we include in addition measurements
reported in the two recent papers~\cite{snonc,snodaynight} by the
Sudbury Neutrino Observatory (SNO) collaboration.

What are the predicted values of the ten most informative
quantities that can be measured in the reasonably near future in
solar neutrino experiments?  We use in
section~\ref{sec:predictions} the allowed regions in neutrino
parameter space (obtained in section~\ref{sec:global}) to predict
in table~\ref{tab:afterpredictions} the expected range of the most
promising quantities that can be measured accurately in the
BOREXINO ~\cite{borexino} and KamLAND~\cite{kamland} ${\rm ^7Be}$
solar neutrino experiments and in the KamLAND reactor experiment,
as well as the spectrum distortion and the day-night asymmetry in
the SNO CC measurements. We also include predictions for a generic
$p-p$ neutrino-electron scattering detector. To assess the
robustness of the predictions, we compare the values predicted
using the ``After April 20, 2002'' global solution
(table~\ref{tab:afterpredictions}) with the values predicted using
the ``Before April 20, 2002'' solution
(table~\ref{tab:beforepredictions}).

We summarize and discuss our conclusions in
section~\ref{sec:discuss}\footnote{Several
papers~\cite{bargernc,strumianc,indiannc} have appeared
essentially contemporaneously with the present paper and treat
some of the same topics with somewhat similar results, although
refs.~\cite{bargernc,strumianc,indiannc}  have not calculated
predictions for the ten future solar neutrino observables studied
in the present paper. On a technical level, as far as we can tell,
these papers have not included the correlations and the energy
dependences of the neutrino absorption cross sections for the
chlorine and gallium experiments (see the Appendix of
ref.~\cite{sterile}). Also, the potential effects of distortions
on the interpretation of the SNO data in terms of individual rates
and their error correlations (see
section~\ref{subsec:aficionados}) were not treated in detail in
the originally-posted versions, although more complete treatments
have been made in later versions~\cite{bargernc}.
 Both of these effects are
included in the present paper. A concise but insightful and
informative discussion of the effects of the recent SNO
measurements on solar neutrino oscillations is given in the
original SNO NC paper~\cite{snonc}. The interested reader may wish
to consult in addition a number of recent papers,
refs.~\cite{bgp,krastev01,foglipostsno,barger2001,bayesian,goswami},
that have determined from a variety of perspectives the allowed
solar neutrino oscillation solutions following the June, 2001
announcement of the SNO CC measurement~\cite{sno2001}.}.

We do not discuss in this paper the implications of the agreement
between the measured~\cite{snonc} flux of active $^8$B solar
neutrinos and the predicted~\cite{bp2000} standard solar model
$^8$B neutrino flux. The agreement is accidentally too good to be
true [see eq.~(\ref{eq:fbrangesno})]. As more measurements are made
of the neutrino flux and of the solar model parameters the
agreement should become less precise. We are aware of several
recent and ongoing  measurements, which are currently not in good
agreement, for the low energy cross section factor $S_{17}$, to
which the calculated standard solar model $^8$B neutrino flux is
proportional. Until the new laboratory measurements of $S_{17}$
converge to a better defined range, we continue to use the
standard value adopted in BP00\footnote{For an insightful
discussion of the predicted and measured total $^8$B neutrino
flux, see ref.~\cite{bargernc}.}.

\section{Global oscillation solutions}
\label{sec:global}

We describe in section~\ref{subsec:strategies} the oscillation
solutions that are allowed with three different analysis
prescriptions (see figure~\ref{fig:global3}). We have at different
times used all three of the analysis strategies and various
colleagues have advocated strongly one or the other of the
strategies described here.  Given the recent SNO NC measurement,
we now prefer the strategy in which the $^8$B neutrino flux is
treated as a free parameter. This strategy is implemented in
figure~\ref{fig:global3}a (see the discussion below). A comparison
of the results obtained using the three strategies allows one to
test the robustness of any conclusion to the method of analysis.

We present in section~\ref{subsec:allowed} and
table~\ref{tab:bestfitsa}
 the best-fit oscillation parameters
for the allowed and the disfavored solutions, treating the  $^8$B
neutrino flux as a free parameter. We discuss in this section the
CL at which different oscillation solutions are acceptable.

In section~\ref{subsec:allowedranges}, we present and discuss the
allowed ranges for $\Delta m^2$, $\tan^2 \theta$, and the total
active flux of $^8$B solar neutrinos.

We describe in section~\ref{subsec:predicteddn} and in
figure~\ref{fig:pee} the predicted dependence of the survival
probability as a function of energy and of  day or night for the
best-fit LMA, LOW, and vacuum solutions. Figure~\ref{fig:pee} in
particular provides a succinct overview of the energy and
day-night dependences of the best-fit survival probabilities.

 The SNO experiment detects
CC, NC, ES ($\nu-e$ scattering), and background events. The rates
from these different processes are correlated because they are,
with the present data, observed most accurately in a mode in which
all of the events are considered together and a simultaneous
solution is made for each of the separate processes using their
known angular dependences (with respect to the solar direction) and
their radial dependences in the detector, as well as information
from direct measurements of the background
~\cite{snonc,snodaynight,snourl}. Details of how this analysis is
done are given in refs.~\cite{snodaynight,snourl}.

Since the different processes are coupled together in the
analysis, the inferred values for the measured fluxes in each of
the CC, NC, and ES modes depend upon the assumed distortion of the
CC and ES recoil energy spectra, which in turn depend upon the
assumed $\Delta m^2$ and $\tan^2 \theta$.  In order to avoid this
cycle, the SNO collaboration presented~\cite{snodaynight} results
for the CC, NC, and ES fluxes that were determined by assuming
that the CC and ES recoil energy spectra are undistorted by
neutrino oscillations or any other new physics.

The fluxes inferred using the hypothesis of undistorted energy
spectra has been used by ref.~\cite{indiannc}
in their analyses of the SNO data. This approximation is excellent
for the LMA and LOW solutions, as we shall show later in this
section (see, e.g., results quoted in
section~\ref{subsec:allowed}), because for these solutions the
expected distortions are indeed small. The approximation is less
accurate for other solutions in which the distortions are more
significant.

In order to obtain the results presented in the following
subsections, we used a two-step procedure to take account,  for
each value of $\Delta m^2$ and $\tan^2 \theta$, of the energy
dependence of the survival probability.  We have used in these
calculations experimental data provided by the SNO
collaboration~\cite{snourl}. We describe how we have carried out
the details of this analysis in section~\ref{subsec:aficionados}.
The results obtained using the full SNO day-night energy spectrum
are presented in section~\ref{sec:fullsno}.

\FIGURE[!t]{
\centerline{\psfig{figure=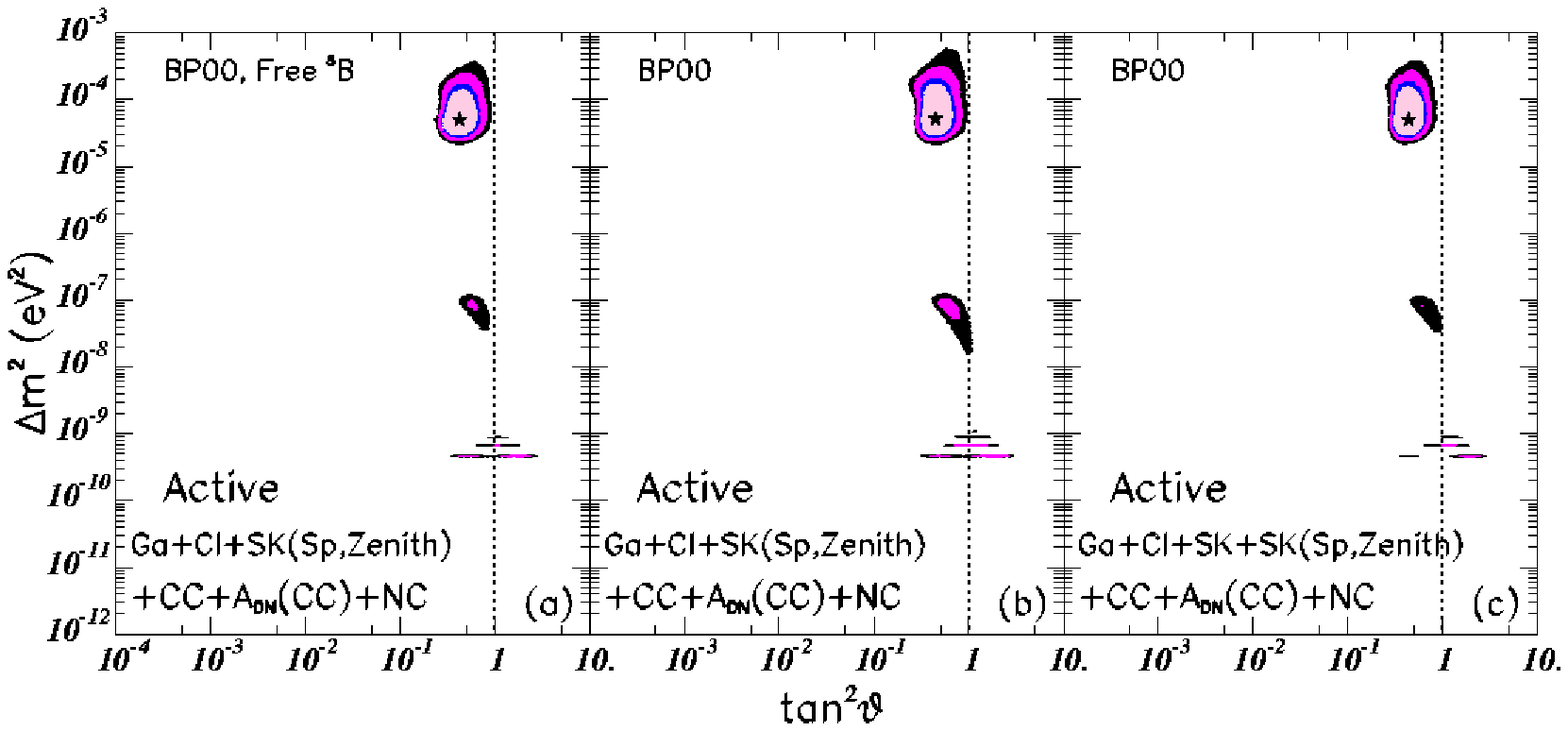,width=5.5in}}
\caption{{\bf Global neutrino oscillation solutions for three
different analysis strategies.}  The strategy used in constructing
panel (a) treats the $^8$B solar neutrino flux as a free parameter
to be determined by the experimental data together with $\Delta
m^2$ and $\tan^2\theta$. The strategies corresponding to panels
(b) and (c) include the theoretical uncertainty in the $^8$B
neutrino flux, but differ in how they treat the total rate
measured in the Super-Kamiokande experiment. The input data used
in constructing figure~\ref{fig:global3} include the neutrino
fluxes and uncertainties predicted by the BP00 solar
model~\cite{bp2000} and the total measured  CC and NC event rates
from the SNO experiment ~\cite{snonc}, the SNO day-night
asymmetry~\cite{snodaynight}, the Chlorine~\cite{chlorine} and
Gallium (averaged)~\cite{sage2002,gallex,gno,sage} event rates, as
well as the zenith angle-recoil energy spectrum data presented by
Super-Kamiokande~\cite{superk}. The rates from the GALLEX/GNO and
SAGE experiments have been averaged to provide a unique data
point ($72.4 \pm 4.7$ SNU). The CL contours shown in the figure
are $90$\%, $95$\%, $99$\%, and $99.73$\% ($3\sigma$). The global
best-fit points are marked by a star. \label{fig:global3}}}

\subsection{Three strategies} \label{subsec:strategies}

Figure~\ref{fig:global3} shows the allowed ranges of the neutrino
oscillation parameters, $\Delta m^2$ and $\tan^2\theta$, that were
computed using the three different analysis approaches that have
been used previously in the literature.  We use the analysis
methods and procedures described in
refs.~\cite{robust,sterile,bks2001,cc2001,bks98,bgp,fogli}), see
especially section~3.3 of ref.~\cite{robust} and the Appendix of
ref.~\cite{sterile}.  We follow refs.~\cite{lisitan,GFM} in using
$\tan^2\theta$ (rather than $\sin^2 2\theta$) in order to display
conveniently the solutions on both sides of $\theta = \pi/4$.

Figure~\ref{fig:global3}a presents the result for our standard
analysis, (a), the free ${\rm ^8B}$ analysis. The strategy used in
this standard analysis takes account of the BP00 predicted fluxes
and uncertainties for all neutrino sources except for $^8$B
neutrinos. The normalization of the $^8$B neutrino flux is treated
as a free parameter for analysis strategy (a) but is constrained
by the BP00 prediction and error for strategies (b) and (c).
Analysis strategy (a) considers all the experimental data except
for the Super-Kamiokande total event rate. The recoil electron
zenith angle-recoil energy spectrum data represent the
Super-Kamiokande total rate in this approach. Details of the
treatment of the Super-Kamiokande errors are given in
Sec.~\ref{subsec:aficionadosFF}.

Figure~\ref{fig:global3}b displays the results of a calculation,
analysis strategy (b), which is the same as for the standard case,
figure~\ref{fig:global3}a, except that the ${\rm ^8B}$ neutrino
flux is  constrained by the BP00 solar model prediction. Prior to
the SNO measurement of the NC flux~\cite{snonc}, (b) was our
standard analysis strategy (cf. ref.~\cite{robust}). Following the
SNO NC measurement, we prefer to use the neutrino data to
determine the flux normalization and therefore to test, rather
than assume, the standard solar model prediction for the $^8$B
neutrino flux.

Figure~\ref{fig:global3}c was constructed by an analysis similar
to that used to construct figure~\ref{fig:global3}b except that
for figure~\ref{fig:global3}b the total Super-Kamiokande rate is
included explicitly together with a free normalization factor for
the zenith angle-recoil energy spectrum of the recoil electrons.
This procedure has been used especially effectively by the Bari
group~\cite{foglipostsno}\footnote{The strategies (a), (b), and
(c) described here correspond, respectively, to the strategies
(c), (a), and (b) discussed in detail in ref.~\cite{robust}.}

 In all cases, we have accounted for the energy and
time dependence of the Super-Kamiokande data by using their zenith
angle-recoil energy spectrum data.  The available Super-Kamiokande
zenith angle-recoil energy spectrum consists of 44 data points,
corresponding to six night bins and one day bin for six energy
bins between 5.5 and 16 MeV electron recoil energy, plus two daily
averaged points for the lowest ($5.0<E<5.5$ MeV) and the
highest ($E>16$ MeV)  energy bins. Alternatively, one could use
their day-night energy spectra as given in 19 energy bins each for
the day and for the night periods. Using the more complete zenith
angle-recoil energy spectrum data allows for a better
discrimination between the LMA and the LOW solutions. Within the
LMA regime, oscillations in the Earth are rapid and therefore LMA
predicts a rather flat distribution in zenith angle. On the other
hand , LOW corresponds to the matter dominated regime of
oscillations in the Earth and LOW predicts a well defined
structure of peaks in the zenith distribution \cite{baltz,alexei}.
The non-observation of such peaks in the zenith angle-recoil
energy spectrum data decreases the likelihood of the upper part of
the LOW solution as compared to the LMA solution. If one were to
use instead the day-night spectra with only night average data,
this feature would be missed.

The main difference in the allowed regions shown in the three
panels of figure~\ref{fig:global3} is that strategy (b) allows a
slightly larger region for the LOW solution. For the
marginally-allowed (or disallowed) LOW region, the required value
for the $^8$B neutrino flux can be significantly different from
the standard solar model or the SNO NC value (the two are virtually
indistinguishable). Thus the inclusion of the  standard solar
model uncertainty for the $^8$B neutrino flux increases the error
used in computing the $\chi^2$ for  this strategy, which has the
effect of enlarging the allowed region.

In constructing figure~\ref{fig:global3}, we assumed that only
active neutrinos exist.  We derive therefore the allowed regions
in $\chi^2$ using only two free parameters: $\Delta m^2$ and
$\tan^2\theta$\footnote{The allowed regions for a given CL are
defined in this paper as the set of points satisfying the
condition $$    \chi^2 (\Delta m^2,\theta)
    -\chi^2_{\rm min}\leq \Delta\chi^2 \mbox{(CL, 2~d.o.f.)} ,
$$ with $\Delta\chi^2(\mbox{CL, 2~d.o.f.}) = 4.61$, $5.99$,
$9.21$, and $11.83$ for CL~= 90\%, 95\%, 99\% and 99.73\%
 ($3\sigma$)}. We use the standard least-square analysis
approximation for the definition of the allowed regions with a
given confidence level. As shown in ref.~\cite{bayesian} the
allowed regions obtained in this way are very similar to those
obtained by a Bayesian analysis.

\subsection{Allowed and disfavored solutions}
\label{subsec:allowed}

Table~\ref{tab:bestfitsa} gives for our standard analysis strategy
 (cf. figure~\ref{fig:global3}a) the best-fit values for $\Delta
m^2$ and $\tan^2 \theta$ for all the neutrino oscillation
solutions that were discussed in our previous analysis in
ref.~\cite{bgp}. The table also lists the values of $\chi_{\rm
min}^2$ for each solution. The regions for which the local value
of $\chi_{\rm min}^2$ exceeds the global minimum  by more than
$11.83$ are not allowed at $3\sigma$ CL The number of degrees of
freedom in this analysis is 46: 44 (Super-Kamiokande zenith-angle
energy spectrum) + 2 (Ga and Cl rates) + 3 ( SNO CC rate , SNO NC
rate and $A_{DN}$(SNO CC) ) $-$3 parameters ( $\Delta {\rm m}^2$,
$\theta$, and $f_{\rm B}$).

\TABLE[!t]{\caption{\label{tab:bestfitsa} {\bf Best-fit global
oscillation parameters with all solar neutrino data.}  The table
gives for the the best-fit values for $\Delta m^2$, $\tan^2
\theta$, $\chi^2_{\rm min}$, and g.o.f. for all the oscillation
solutions among active solar neutrinos that have been previously
discussed (see, e.g., ref.~\cite{bgp}). The quantity $f_{\rm B}$
measures the $^8$B solar neutrino flux in units of the predicted
BP00 neutrino flux, see  eq.~(\ref{eq:fbdefn}). The oscillation
solutions are obtained by varying the $^8$B flux as free parameter
in a consistent way: simultaneously in the rates and in the night
and day spectrum fits. The differences of the squared masses are
given in ${\rm eV^2}$. The number of degrees of freedom is 46 [44
(zenith spectrum) + 4 (rates) + 1 ($A_{DN} (CC)$) $-$3
(parameters: $\Delta {\rm m}^2$, $\theta$, and $f_{\rm B}$)]. The
goodness-of-fit given in the last column is calculated relative to
the minimum for each solution. (Solutions that have $\chi_{\rm
min}^2 \geq 45.2 + 11.8=57.0$ are not allowed at the $3\sigma$
CL)}
\begin{tabular}{lccccc}
\noalign{\bigskip} \hline \noalign{\smallskip} Solution&$\Delta
m^2$&$\tan^2 (\theta)$& $f_{\rm B,best}$
& $\chi^2_{\rm min}$ & g.o.f. \\
\noalign{\smallskip} \hline \noalign{\smallskip} LMA&
$5.0\times10^{-5}$  &$4.2\times10^{-1}$& 1.07
& 45.5 &$49$\% \\
LOW& $7.9\times10^{-8}$  &$6.1\times10^{-1}$& 0.91
& 54.3 &$19$\%\\
VAC& $4.6\times10^{-10}$  &$1.8\times10^{0}$& 0.77
& 52.0 &$25$\%\\
SMA& $5.0\times10^{-6}$  &$1.5\times10^{-3}$& 0.89
& 62.7 &$5.1$\%\\
Just So$^2$ & $5.8\times10^{-12}$  & $1.0\times10^{0} $& 0.46
& 86.3  &$\sim 0$\%\\
Sterile VAC & $4.6\times10^{-10}$ & $2.3\times10^{0}$ & 0.81
& 81.6 & $\sim 0$\%\\
Sterile Just So$^2$
& $5.8\times10^{-12}$  &\hskip -6pt$1.0\times10^{0} $ & 0.46
& 87.1 & $\sim 0$\%\\
Sterile SMA & $3.7\times10^{-6}$ & $4.7\times10^{-4}$ & 0.55
& 89.3 & $\sim 0$\%\\
\noalign{\smallskip} \hline
\end{tabular}
}

Within the MSW regime, only the LMA and LOW solutions are allowed
at $3\sigma$ with the currently available data. The difference in
$\Delta \chi^2$ between the global best-fit point (in the LMA
allowed region) and the best-fit LOW point is  $\Delta \chi^2 =
8.8$, which implies that the LOW solution is allowed only at the
{$98.8$\% CL ($2.5\sigma$).

The vacuum solution is currently allowed at the $96$\% CL
($2.1\sigma$). This solution is also allowed in our Before
analysis, which we present later in figure~\ref{fig:beforeafter},
at a CL better than 95\% CL (and also in the most recent
Super-Kamiokande analysis \cite{smy2002}). But it was not found by
the SNO collaboration at the 3$\sigma$ level. In the footnote that
appears in section~\ref{sec:discuss}, we provide a possible
explanation for the absence of allowed vacuum solutions in the SNO
analysis.

For oscillation solutions for which the survival probability does
not depend strongly upon  energy, such as the LMA and LOW
solutions, the effects of including~\cite{snodaynight,snourl}  the
energy distortion in the determination of the rates is very small.
We find, using the procedure described in
section~\ref{subsec:aficionados}, that the central values of the
CC and NC fitted rates for the best fit points in LMA (LOW) in
table~\ref{tab:bestfitsa} are shifted by $+0.5$\% and
$-1.5$\% ($0$\% and $+3\%$), respectively, with respect to the
values obtained under the hypothesis of no energy distortion. This
results in an increase of the $\Delta\chi^2$ between LMA and LOW
of $\sim 0.2$ (i.e., by about $2$\%). For solutions with stronger
energy dependences such as SMA (VAC), the effects are larger and
lead to shifts in the central values of the CC and NC rates of
$-1.5$\% and $+15.5$\% (+10\% and -11\%).

We have made a number  of checks on the stability of our
conclusions regarding the global solutions. The $\Delta\chi^2$
between LMA and LOW is rather robust under small changes in the
method of error treatment and in the fitting procedure. However,
the CL  at which the VAC and the QVO solutions (the region
between the LOW solutions and the VAC solutions, see the
insightful discussions by Friedland~\cite{qvo} and Lisi et
al.~\cite{qvo2}) are allowed may fluctuate from just below to
just above the $3\sigma$ limit, depending upon details of the
analysis. For example, all VAC solutions are disfavored at
$3\sigma$ if one ignores the anti-correlation between the
statistical errors of the NC and the CC rates.

The best-fitting pure sterile solution is  VAC, which is excluded
at  $5.4\sigma$ CL (for $3$d.o.f.). Before April 20, 2002,
the best-fitting pure sterile solution was SMA sterile, which was
acceptable at $3.6\sigma$.

Oscillations into an admixture of active and sterile neutrinos are
still possible as long as the assumed total $^8$B neutrino flux is
increased appropriately~\cite{sterile,bargernc}. This uncertainty
can be reduced by combining SNO solar neutrino data with results
from the terrestrial KamLAND reactor experiment~\cite{sterile}.

Our value for
$\Delta \chi^2 = \chi^2_{\rm LOW} -\chi^2_{\rm LMA}$ may be lower
than most other groups who do similar calculations because we
include the effect of the correlations and the energy dependence
of the neutrino absorption cross sections for the gallium and
chlorine solar neutrino experiments (see the Appendix of
ref.~\cite{sterile}). The cross section effects correspond to
$\Delta \chi^2 = -1.3$ for the current global allowed solution,
strategy (a).  Taking this effect into account, our values for
$\Delta \chi^2$ between the LMA and LOW solutions seems to be in
general agreement with most other
authors~\cite{snonc,bargernc,strumianc,indiannc}.

\subsection{Allowed ranges of mass, mixing angle, and $^8$B
neutrino flux} \label{subsec:allowedranges}

The upper limit on on the allowed value of $\Delta m^2$ is
important for neutrino oscillation experiments, as stressed in
ref.~\cite{krastev01}. In units of ${\rm eV^2}$, we find for the
LMA solution the following $3\sigma$ limits on $\Delta m^2$
\begin{equation}
2.3\times 10^{-5} < \Delta m^2 < 3.7 \times 10^{-4}.
\label{eq:masslimitlma}
\end{equation}
The LMA solar neutrino region does not reach the upper bound for
$\Delta m^2$ imposed by the CHOOZ reactor data \cite{chooz}, i.e.,
$\Delta m^2\leq 8\times 10^{-4}$ ${\rm  eV^2}$. Prior to April 20,
2002, the solar neutrino data alone were not sufficient to exclude
LMA masses above the CHOOZ bound.

For the LOW solution only the following small mass range is
allowed,
\begin{equation}
3.5\times 10^{-8} < \Delta m^2 < 1.2 \times 10^{-7}.
\label{eq:masslimitlow}
\end{equation}

Many authors (see, e.g., ref.~\cite{bimaximal} and references
quoted therein) have discussed the possibility of bi-maximal
neutrino oscillations, which in the present context implies
$\tan^2 \theta = 1$. Figure~\ref{fig:global3} shows that precise
bi-maximal mixing is disfavored for both the  LMA  and LOW
solutions. Quantitatively, we find that there there are no
solutions with $\tan^2 \theta = 1$ at the $3.3\sigma$ CL for the
LMA solution, at the $3.2\sigma$ CL for the LOW solution, and at
the $2.8\sigma$ CL for the VAC solutions. These results refer to
our standard analysis strategy, corresponding to panel (a) of
figure~\ref{fig:global3}.

Of course, approximate bi-maximal mixing is now heavily favored.
Atmospheric neutrinos oscillate with a large mixing
angle~\cite{atmospheric} and all the currently allowed solar
oscillations correspond to large mixing angles (see
figure~\ref{fig:global3}).

How close are the solar neutrino mixing angles to $\pi/4$? At
three sigma, we find the following allowed range for the LMA
mixing angle
\begin{equation}
 0.24 < \tan^2 \theta < 0.89,
 \label{eq:tanlimitlma}
 \end{equation}
and for the LOW solution
\begin{equation}
 0.43< \tan^2 \theta < 0.86.
 \label{eq:tanlimitlow}
 \end{equation}

Let $f_{\rm B}$ be the ${\rm ^8B}$ neutrino flux inferred from
global fits to all the available solar neutrino data. Moreover,
let $f_{\rm B}$ be measured in units of the best-estimate
predicted BP00 neutrino flux ($5.05\times 10^6{\rm
\,cm^{-2}s^{-1}}$),
\begin{equation}
f_{\rm B} ~ =~ \frac{\phi({\rm ^8B})}{\phi({\rm ^8B})_{\rm BP00}}.
\label{eq:fbdefn}
\end{equation}
The best-fit values for $f_{\rm B}$ for each of the oscillation
solutions are listed in the fourth column of
table~\ref{tab:bestfitsa}.

The value of $f_{\rm B}$ found by the SNO collaboration from their neutral
current measurement via a simultaneous solution for all reactions
in the SNO detector is~\cite{snonc}
\begin{equation}
 f_{\rm B} ~=~ 1.01[1 \pm 0.12]~(1\sigma).
\label{eq:fbrangesno}
\end{equation}
For the global solution shown in figure~\ref{fig:global3}a, the
$1\sigma[3\sigma]$ allowed range of $f_{\rm B}$ in the LMA
solution region is
\begin{equation}
 f_{\rm B} ~=~ 1.07\pm 0.08~({1\sigma})~~[f_{\rm B} ~=~
 1.07^{+0.23}_{-0.25}]~({3\sigma})~({\rm LMA}).
\label{eq:fbrangelma}
\end{equation}
The $3\sigma$ range of $f_{\rm B}$ in the LOW solution region is
\begin{equation}
 f_{\rm B} ~=~  0.91^{+0.03}_{-0.02}~(3\sigma ,\,{\rm LOW}).
\label{eq:fbrangelow}
\end{equation}
For both eq.~(\ref{eq:fbrangelma}) and eq.~(\ref{eq:fbrangelow}),
the range of $f_{\rm B}$ was calculated by marginalizing over the
full space of oscillation parameters ($\Delta m^2$, $\tan^2
2\theta$) using the global minimum value for $\chi^2$ which lies
in the LMA allowed region.

 The allowed range of the $^8$B active solar
neutrino flux derived from the global oscillation solution and
given in eq.~(\ref{eq:fbrangelma}) is slightly more restrictive
than was found by the SNO collaboration using just their NC
measurement (cf. eq.~\ref{eq:fbrangesno}).

\FIGURE[h!t]{ \centerline{\psfig{figure=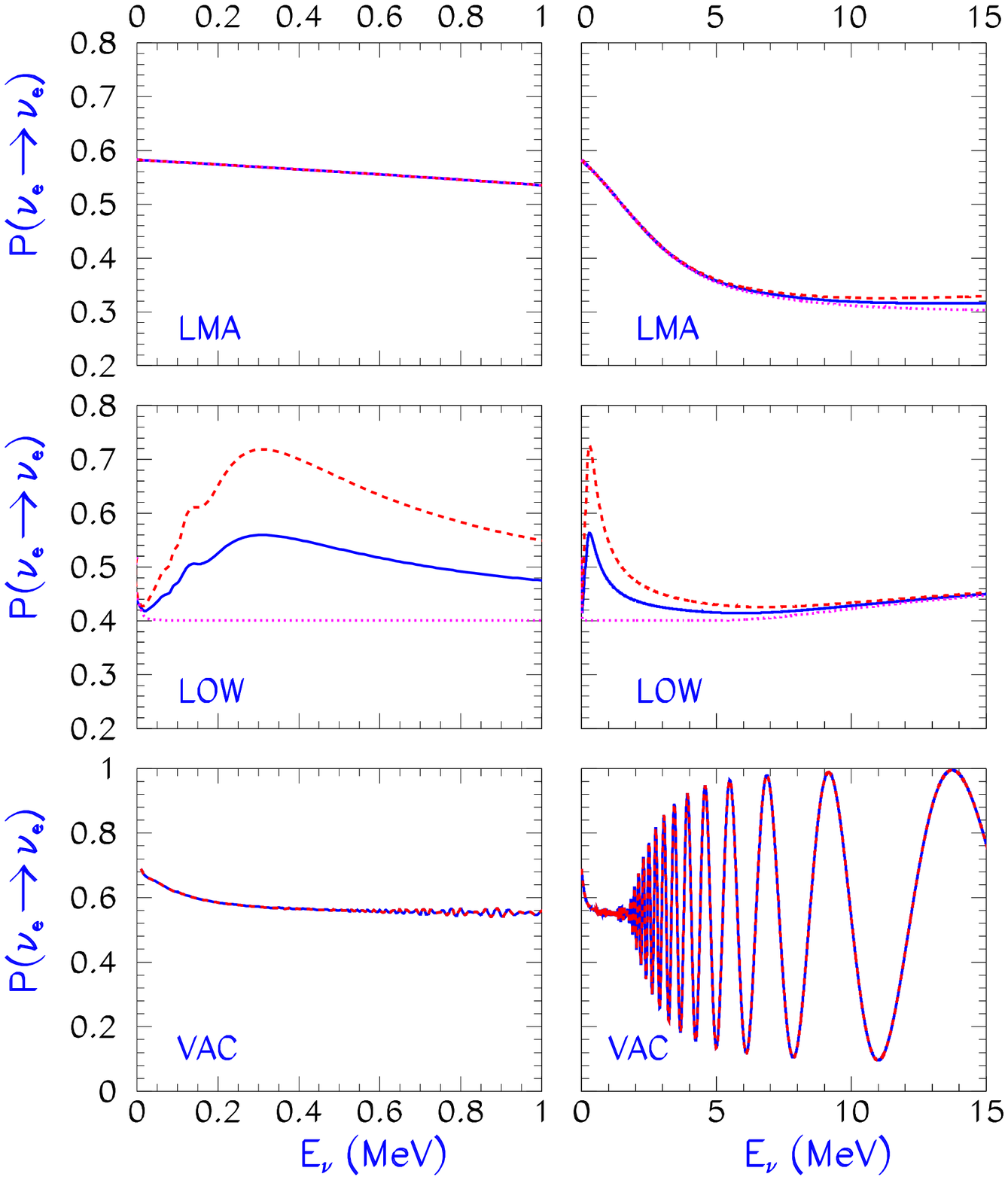,width=4in}}
\caption{{\bf Survival Probabilities.} The figure presents the
yearly-averaged best-fit survival probabilities determined with
strategy (a) for an electron type neutrino that is created in the
center of the Sun and arrives at a detector on Earth. The full
line refers to the average survival probabilities computed taking
into account regeneration in the Earth and the dotted line refers
to calculations for the daytime that do not include earth
regeneration. The dashed line include regeneration at night.  The
regeneration effects are computed for the location of the SNO
detector for the right hand panels and for the location of the
Gran Sasso Underground Laboratory for the left hand panels. There
are only slight differences between the computed regeneration
effects for detectors located at the positions of SNO,
Super-Kamiokande, and Gran Sasso (see ref.~\cite{bkdaynight}). The
LOW solutions in the right-hand panel are averaged over a small
energy band, $0.1$ MeV, to suppress rapid oscillations caused by a
sensitive dependence upon the Earth-Sun distance. The vacuum
solutions are averaged over an energy band of $\pm 0.05$ MeV.
\label{fig:pee}}}

\subsection{The Predicted Energy and Day-Night Dependence}
\label{subsec:predicteddn}

Figure~\ref{fig:pee} shows the predicted energy dependence and the
day-night asymmetry for the survival probability $P (\nu_e
\rightarrow \nu_e)$ of the allowed MSW oscillation solutions, the
LMA and LOW solutions, as well as the best-fit vacuum solution.

We plot in  figure~\ref{fig:pee} the best-fit survival
probabilities for an electron neutrino that is created in the Sun
to remain an electron neutrino upon arrival at a terrestrial
detector. The right-hand panels present the survival probability
for energies between $0$ and $15$ MeV. The left-hand panels
present blow-ups of the behavior of the solutions at energies less
than $1$ MeV.

The energy dependence of both MSW solutions is predicted to be
very modest above $5$ MeV, in agreement with the fact that no
statistically significant distortion of the recoil energy spectrum
has yet been observed in the Super-Kamiokande
experiment~\cite{superk}. For the SNO CC measurements,
table~\ref{tab:afterpredictions} shows that the expected
distortions of the first and second moments of the electron recoil
energy spectrum are predicted to be too small to be measurable
with statistical significance.

At energies below $5$ MeV, both the LMA and the LOW solutions are
predicted to exhibit a significant energy dependence.

At the energies at which water Cherenkov experiments have been
possible, the survival probability $P (\nu_e \rightarrow \nu_e)
\approx 1/3$, i.e., approximately one over the number of known
neutrinos. According to MSW theory, this is an accident. We see
from figure~\ref{fig:pee} that a survival probability of order
$0.5$ is predicted for energies less than $1$ MeV.

The day-night asymmetry is potentially detectable for the LMA
solution only at energies above $5$ MeV; the predicted asymmetry
increases with energy (see figure~\ref{fig:pee}). The situation is
just the opposite for the LOW solution. The day-night asymmetry is
large below $5$ MeV and relatively small above $5$ MeV.

The vacuum solution has rapid oscillations in energy above $2$ MeV
 (see figure~\ref{fig:pee}), but these oscillations would be very
difficult to observe because they tend to average out over a
typical experimental energy bin. At lower energies, the vacuum
oscillations will also appear to be smooth, again because of the
difficult of resolving in energy the oscillations.

\subsection{Analysis details for aficionados: the two-step procedure}
\label{subsec:aficionados}

The relatively precise values of the CC and NC rates given in the
recent  SNO paper~\cite{snonc}  were extracted by a fit to the
observed recoil  energy spectrum using the assumption of an
undistorted spectrum, {\sl i.e.} for a survival probability that
is constant in energy\footnote{They also quote the NC rate
obtained from a fit with arbitrary distortion, which results into
a much less precise determination.}. These rates cannot be used
directly to test an oscillation hypothesis which would cause
significant distortions of the CC and ES recoil energy spectra.
Since they were aware of the inconsistency in doing so, the SNO
collaboration did not use in their oscillation analysis the rates
extracted assuming undistorted spectra.
Instead, they correctly
performed a direct fit to their summed spectrum, including the
three contributions from NC, CC and ES (as well as the background)
computed for each point in oscillation parameter space.

We include the spectral distortions in a somewhat different way
than was done by the SNO collaboration. We perform a two step
analysis, which has advantages that we discuss at the end of this
section. We use the data generously provided by the SNO
collaboration~\cite{snourl}.
\begin{itemize}
\item First, for a given point in neutrino oscillation parameter space, we
compute for each of the $17$ SNO energy bins the oscillation
probabilities that are appropriate for evaluating the CC and ES
recoil energy spectra.  We also compute the number of events that
would be expected in each bin for an undistorted energy spectrum
 (survival probabilities equal to unity everywhere).  We multiply
the number of events in a given energy bin for the undistorted
spectrum by the oscillation probabilities and then add the CC and ES
energy spectrum  to the NC energy spectrum
 (which is undistorted). We fit the observed SNO energy spectrum with
 an arbitrary linear combination of the three energy spectra, CC, ES, and NC,
 plus the background. The coefficients of the three spectra are
 evaluated by minimizing the $\chi^2$ fit to the observed SNO spectrum.
  In order to test easily whether the individual results are physically plausible,
 we normalize the coefficients so that they are unity for an undistorted
 spectrum. We have verified that the inferred NC, CC,  and ES rates are always
within the expected physical range. We never encounter unphysical
situations: the extracted rates for all three processes, NC, CC,
and ES, are always positive and the CC rate never exceeds the NC
rate in any part of the parameter space,  even in those regions
with large spectral distortions\footnote{In
section~\ref{subsec:allowed}, we noted that the normalized
 coefficients differ from unity by typically $1$\% or $2$\% for the LMA
 and LOW solutions, i. e., the spectra distortion is not important for
 these cases. For the SMA and VAC solutions, which have stronger energy
 dependences for the survival probabilities, the shifts can be $\sim
 10$\%.}.
 In this way,  we
obtain as a function of the oscillation parameters, the best-fit
CC and NC  measured rates
 (and their corresponding statistical errors, which are strongly
anti-correlated).
\item Second, with these NC and CC rates, we compute the $\chi^2$
for each particular point in oscillation space, combining these
two measurements with the available results of all other solar
neutrino experiments. We use the three different analysis
strategies discussed in ref.~\cite{robust} and in
section~\ref{subsec:strategies} of this paper.  We use for the SNO
day-night asymmetry the result quoted for an undistorted
spectrum~\cite{snodaynight}, since the small effect of the
distortion is expected to nearly cancel out in the asymmetry. In
combining the NC and CC data, we take account of the correlation
between their errors (both statistical and systematic). We make
the approximation that the systematic uncertainties have the same
percentage values~\cite{snonc,snodaynight} as they have for the
undistorted rates. This approximation is also used, among others,
by the SNO collaboration in their recent
analysis~\cite{snodaynight}. In their recent work, Fogli et.al
\cite{foglinc} discuss the possible effect of this approximation
and find that the approximation is accurate near the local minima
but may induce some inaccuracy in the allowed ranges near the
boundaries of the $3\sigma$ limits. The results of our full-SNO
analysis, presented in section.~\ref{sec:fullsno}, confirm the
conclusion of Fogli et al. We find essentially the same effects as
reported by Fogli et al.
when we no longer make the assumption that the uncertainties have
the same percentage values as are computed for a undistorted
spectrum.

The constraints on the relations between the ES, CC, and NC rates
that exist for active oscillations are implemented in step two
after the CC and NC extracted rates are included in the global fit
together with the ES spectra from Super-Kamiokande. We do not
include the SNO ES rate in our global fit, since the
Super-Kamiokande ES rate is much more precise, but we do  verify,
that the extracted SNO ES rate is always compatible with the
Super-Kamiokande ES rate.

\end{itemize}

What are the advantages of the two-step procedure? The most
obvious advantage is speed. It is faster to solve for the allowed
neutrino oscillation parameters in $\chi^2$ space using just the
three SNO data points  representing the CC and NC fluxes  and the
day-night difference  than it
is to solve for the allowed parameters using all $34$ points in
the SNO day-night spectrum. Moreover, the use of two distinct
methods, the two-step procedure described here and the full-SNO
procedure utilized in section~\ref{sec:fullsno}, permits a test of
the robustness of the conclusions. In the two-step procedure, the
SNO data are represented by only $3$ data points in the global
$\chi^2$ (compared to two rate parameters for the radiochemical
experiments, chlorine and gallium). Computation of the global
$\chi^2$ for the full-SNO procedure requires the use of $34$ data
points from SNO. Thus for the two-step procedure, SNO is
represented in the global $\chi^2$ by a comparable number of data
points as the radiochemical experiments while in the full-SNO
procedure the SNO experiment contributes more than $10$ times as
many data points as the radiochemical experiments. Both methods,
the two-step procedure and the full-SNO procedure, give (see
section~\ref{subsec:comparison})  similar allowed regions and very
similar predicted values for solar neutrino observables that will
be measured in the future.

Finally, we note that the two-step method has the advantage of
transparency. By looking at the computer which  contains the
coefficients of the NC, CC, and ES energy spectra we can
immediately see the effect of any choice of oscillation parameters
on the different event rates. This visual inspection also allows
us to check quickly for possible unphysical or implausible
solutions.

Given that the spectral energy distortions are strongly
constrained by the fact that
Super-Kamiokande~\cite{smy2002,superk} does not observe a
significant distortion, it was {\it a priori} unlikely that the
two methods of taking account of distortions-the two-step
procedure and the full-SNO procedure-would lead to significantly
different results when analyzing actual solar neutrino
experimental data. This expectation is verified quantitatively in
section~\ref{subsec:comparison}.

\section{Global ``Before and After''}
\label{sec:beforeafter}

\FIGURE[!t]{
\centerline{\psfig{figure=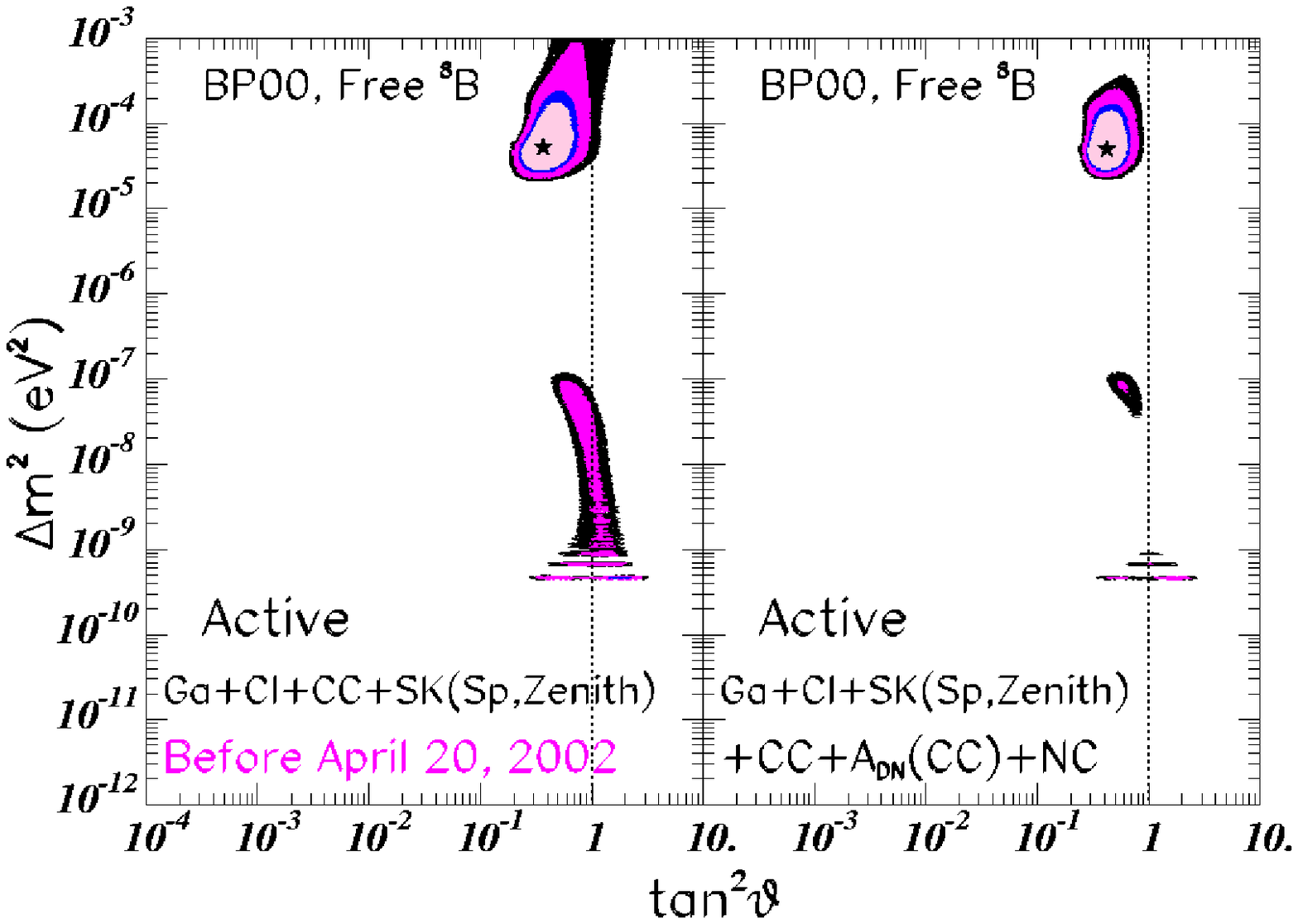,width=4.5in}}
\caption{{\bf Global ``Before and After.''} The left panel shows
the allowed regions for neutrino oscillations computed using solar
neutrino experimental data available prior to April 20 2002. The
right panel shows the allowed region computed with the same
procedure but including the SNO NC and improved CC
data~\cite{snonc} and the SNO day-night
asymmetry~\cite{snodaynight}. In the ``Before'' panel, the LOW
solution is allowed at $97.4$\% and in the ``After'' panel LOW is
allowed at the $98.8$\%. \label{fig:beforeafter}}}

 What is
the impact of the recent SNO measurements~\cite{snonc,snodaynight}
on the globally allowed regions of neutrino oscillation
parameters?

Figure~\ref{fig:beforeafter} compares the allowed regions that are
found with the solar neutrino data available prior to the
presentation of the recent SNO data (i.e., prior to April 20, 2002)
with the allowed regions found including the recent SNO
data~\cite{snonc,snodaynight}. We like to refer to
figure~\ref{fig:beforeafter} as our ``Before and After'' figure.

The left panel of figure~\ref{fig:beforeafter} was computed
including the improvements that we described in
section~\ref{sec:introduction} regarding the neutrino cross
section errors and correlations and the improved average gallium
event rate. Hence figure~\ref{fig:beforeafter}a does not
correspond to any previously published global oscillation
solution, although it could have been computed prior to April 20,
2002.

The recent SNO measurements have greatly shrunk the allowed region
for the LOW solution and have significantly reduced the allowed
region for the LMA solution, as can be seen by comparing the two
panels of figure~\ref{fig:beforeafter}. In particular, maximal
mixing is now not allowed at $3\sigma$ and the region does not
reach the CHOOZ reactor bound, $\Delta m^2\leq 8\times 10^{-4}$
eV$^2$\cite{chooz}.

The LOW solution is now allowed at the $98.8$\% CL Before the
recent SNO measurements, the LOW solution was allowed at $97.4$\%.

\section{Predictions for BOREXINO, KamLAND, SNO and a generic $p-p$ detector}
\label{sec:predictions}

We summarize in this section some of the most important
predictions that follow from the global neutrino oscillation
solutions discussed in section~\ref{fig:global3} and
section~\ref{sec:beforeafter}. To test the robustness of the
predictions given in this section, we have calculated the expected
values and allowed ranges by two different methods: 1) using the
precise values for the CC and NC given in ref.~\cite{snonc}
assuming no spectral energy distortions and 2) taking account of
the potential effects of spectral energy distortions on the SNO
measurements as described in section~\ref{subsec:aficionados}. The
differences in the calculated values and ranges are negligible in
all cases. We present here, for consistency with
figure~\ref{fig:global3} and our previous discussion, the values
calculated taking account of the potential spectral distortions in
the SNO data.

We summarize in section~\ref{subsec:predictionsmsw} the
predictions for the MSW solutions. We answer the following
question in section~\ref{subsec:predictionsvac}: How can we
distinguish between vacuum and MSW solutions?

\subsection{Predictions for MSW solutions}
\label{subsec:predictionsmsw}

Table~\ref{tab:afterpredictions}
presents the best-fit predictions and the expected ranges for
none important future solar neutrino observables. The results
summarized in table~\ref{tab:afterpredictions} correspond to the
right-hand side (the ``After'' panel) of
figure~\ref{fig:beforeafter}. The notation used here is the same
as in ref.~\cite{robust}.

\TABLE[h!!t]{\caption{\label{tab:afterpredictions}{\bf ``After''
Predictions.} This table presents for future solar neutrino
observables the best-fit predictions and $1\sigma$ and $3\sigma$
ranges that were obtained by using analysis strategy (a) and all
solar neutrino data currently available. The best-fit values and
uncertainties given here correspond to the allowed regions in the
right hand panel of figure~\ref{fig:beforeafter}; i.e., this table
was constructed by including the data made available by the SNO
collaboration on April 20, 2002~\cite{snonc,snodaynight}. The
day-night asymmetries for the SNO and the $^7$Be (BOREXINO)
experiments are denoted by $A_{\rm N-D}$ and are defined by
eq.~(\ref{eq:daynightdefn}). The reduced $^7$Be, $p-p$, and $^8$B
event rates are defined by eq.~(\ref{eq:be7reducedrate}) and
eq.~(\ref{eq:ppreducedrate}); the reduced KamLAND CC rate is
defined by eq.~(\ref{eq:kamlandreducedrate}). The first moment of
the recoil electron energy distribution is denoted by $\delta T$
for SNO and $\delta E_{\rm visible}$ for KamLAND; the second
moments are denoted by $\delta \sigma$. The threshold of the
recoil electron kinetic energy used in computing the SNO
observables for this table is $5$ MeV. For the BOREXINO
experiment, we consider electron recoil energies between between
$0.25 ~{\rm MeV}$ and $0.8 ~{\rm MeV}$ (see ref.~\cite{borexino}).
We present the results for the KamLAND reactor observables for two
thresholds, $E_{\rm th}=1.22$ and 2.72 MeV.}

\begin{tabular}{lccc}
\noalign{\bigskip} \hline \noalign{\smallskip}
Observable&  b.f. $\pm 1\sigma$  &LMA  $\pm 3\sigma$  & LOW  $\pm 3\sigma$\\
\noalign{\smallskip} \hline \noalign{\smallskip} A$_{\rm N-D}$ (SNO
CC) (\%) & $5.2^{+3.6}_{-3.5}$ & $5.2^{+9.3}_{-5.2}$ &
$2.7^{+2.7}_{-2.1}$
\\
\noalign{\smallskip}\noalign{\smallskip}$\delta T$ (SNO CC) (\%) &
$-0.17^{+0.19}_{-0.56}$ & $-0.17^{+0.31}_{-1.55}$&
$0.42^{+0.55}_{-0.35}$
\\
\noalign{\smallskip}\noalign{\smallskip}$\delta\sigma$ (SNO CC)
(\%)& $0.03^{+0.17}_{-0.61}$ & $0.03^{+0.25}_{-1.77}$ &
$0.52^{+0.66}_{-0.44}$
\\
\noalign{\smallskip} \noalign{\smallskip} $[$R ($^7$Be)$]$ $\nu-e$
scattering & $0.64 \pm 0.03$ & $0.64^{+0.09}_{-0.05}$ & $0.58\pm
0.05$
\\
\noalign{\smallskip} \noalign{\smallskip} A$_{\rm N-D}$ ($^7$Be)
(\%) & $0.0^{+0.0}_{-0.0}$ & $0.0^{+0.1}_{-0.0}$
& $ 23^{+10}_{-13}$
\\
\noalign{\smallskip} \noalign{\smallskip}$[$R ($^8$B)$]$ $\nu-e$
scattering
& & &\\
($T_{\rm th}=3.5$ MeV) & $0.46\pm 0.03$ & $0.46\pm 0.04$
& $0.46\pm 0.03$ \\
($T_{\rm th}=5$ MeV) & $0.46\pm 0.03$ & $0.46\pm 0.04$
& $0.46\pm 0.03$
 \\
\noalign{\smallskip} \noalign{\smallskip} $[$CC$]$ (KamLAND)
& & &\\
 ($E_{\rm th}=2.72$ MeV) &$0.49^{+0.20}_{-0.17}$
&$0.49^{+0.25}_{-0.26}$ & ---
\\
($E_{\rm th}=1.22$ MeV) &$0.52\pm 0.15$ &$0.52^{+0.20}_{-0.25}$ &
---
\\
\noalign{\smallskip} \noalign{\smallskip} $\delta E_{\rm
visible}$ (KamLAND) (\%)
& & &\\
($E_{\rm th}=2.72$ MeV) & $-7^{+13}_{-2}$ & $-7^{+14}_{-4}$
& ---\\
($E_{\rm th}=1.22$ MeV) & $-9^{+13}_{-3}$ & $-9^{+17}_{-5}$
& ---\\
\noalign{\smallskip} \noalign{\smallskip}
$\delta\sigma$ (KamLAND)(\%)
& & &\\
($E_{\rm th}=2.72$ MeV) & $-5^{+11}_{-10}$ & $-5^{+20}_{-14}$
& ---\\
($E_{\rm th}=1.22$ MeV) & $-8^{+20}_{-8}$ & $-8^{+26}_{-12}$
& ---\\
\noalign{\smallskip} \noalign{\smallskip} $[p$-$p]$ $\nu-e$
scattering
& & &\\
($T_{\rm th}=100$ keV) & $0.705^{+0.031}_{-0.026}$
& $0.705^{+0.073}_{-0.049}$
& $0.683^{+0.035}_{-0.042}$ \\
($T_{\rm th}=50$ keV) & $0.700^{+0.031}_{-0.027}$
& $0.700^{+0.074}_{-0.050}$
& $0.677^{+0.038}_{-0.045}$ \\
\noalign{\smallskip} \hline
\end{tabular}
}

How much of a difference have the recent SNO measurements made in
the expected values for future solar neutrino observables?  The
reader can answer this question by comparing
table~\ref{tab:beforepredictions}  with
table~\ref{tab:afterpredictions} . Here,
table~\ref{tab:beforepredictions} presents the best-fit
predictions and allowed ranges that correspond to using solar
neutrino data available prior to April 20, 20002. In particular,
the values given in  table~\ref{tab:beforepredictions} were
obtained using the global solution shown in the left-hand side,
the ``Before'' panel, of figure~\ref{fig:beforeafter}.

The principal differences between the results summarized in
table~\ref{tab:afterpredictions} and
table~\ref{tab:beforepredictions} are a consequence of the smaller
allowed region of the LOW solution shown in the ``After'' panel of
figure~\ref{fig:beforeafter}. The changes that result from the
modest reduction of the LMA allowed region are generally not very
significant.

 The
predicted day-night asymmetries, $A_{\rm N-D}$, between the
nighttime and the daytime event rates for SNO and for
BOREXINO ($^7$Be $\nu-e$ scattering detector) are given in the
first and third rows, respectively, of
table~\ref{tab:afterpredictions}  and
table~\ref{tab:beforepredictions}. The results presented
correspond to an average over one year. The definition of $A_{\rm
N-D}$ is
\begin{equation}
A_{\rm N-D} ~=~2{\rm  {[Night - Day]} \over {[Night + Day]}}~.
\label{eq:daynightdefn}
\end{equation}
We have used $1$ km steps ($6371$ total points) to compute the
values for the day-night effect in the SNO CC measurement.
Compared to the results obtained using a cruder grid of $50$ km
steps, the best-fit values for the $A_{N-D}$ for the LMA are
shifted up by $\sim 10$\%. The LOW predictions for $A_{N-D}$ are
not significantly affected. The main reason for that shift in the
values calculated for the LMA is the more accurate
parameterization of the PREM density~\cite{prem} of the earth in
the external shells.

The minimum predicted value for the LOW day-night asymmetry,
A$_{\rm N-D}$ ($^7$Be), in BOREXINO is now a whopping-big $10$\%
at $3\sigma$. Prior to April 20, 2002, a $0$\% value of A$_{\rm
N-D}$ ($^7$Be) was allowed even for the LOW solution. There are no
significant differences in the Before and After predictions for
the SNO CC value of A$_{\rm N-D}$.

The first and second moments of the SNO CC spectrum for the case
of no oscillations are $\langle T_0\rangle=7.74 $MeV and $\langle
\sigma_0\rangle=1.87$ MeV. The range of predicted shifts with
respect to these no-oscillation values is always smaller than
$\sim 2$\% within both the LMA and LOW $3\sigma$ regions and is
not significantly different from the range found in the ``Before''
analysis. Larger shifts for the moment second moment ($7$\%) are
possible within the allowed VAC oscillation region. The
non-statistical uncertainties  in measuring the first and second
moments in SNO have been estimated, prior to the operation of the
experiment,  in ref.~\cite{tencommand} and are, respectively,
about $1$\% and $2$\%.

\TABLE[h!!t] {\centering \caption{{\bf ``Before'' Predictions.} This
table presents for future solar neutrino observables the best-fit
predictions and $1\sigma$ and $3\sigma$ ranges that were obtained
by using analysis strategy (a) and solar neutrino data available
before April 20, 2002. The best-fit and uncertainties given here
correspond to the allowed regions in the left hand panel of
figure~\ref{fig:beforeafter}. The format and procedures used in
constructing this table are the same as used in constructing
table~\ref{tab:afterpredictions} except that here we have not
included the recent SNO data~\cite{snonc,snodaynight}.
\protect\label{tab:beforepredictions}}
\begin{tabular}{lccc}
\noalign{\bigskip}
\hline
\noalign{\smallskip}
Observable&  b.f. $\pm 1\sigma$  &LMA  $\pm 3\sigma$  & LOW  $\pm 3\sigma$\\
\noalign{\smallskip}
\hline
\noalign{\smallskip}
A$_{\rm N-D}$ (SNO CC) (\%)
& $4.9^{+4.1}_{-3.4}$
& $4.9^{+11.7}_{-4.9}$
& $1.3^{+3.9}_{-1.3}$
\\
\noalign{\smallskip}\noalign{\smallskip}$\delta T$ (SNO CC) (\%)
& $-0.23^{+0.25}_{-0.57}$
& $-0.23^{+0.39}_{-1.58}$
& $0.25^{+0.71}_{-0.44}$
\\
\noalign{\smallskip}\noalign{\smallskip}$\delta\sigma$ (SNO CC) (\%)
& $0.16^{+0.12}_{-0.78}$
& $0.16^{+0.13}_{-2.0}$
& $0.25^{+0.96}_{-1.02}$
\\
\noalign{\smallskip} \noalign{\smallskip} $[$R ($^7$Be)$]$ $\nu-e$
scattering & $0.66\pm 0.04$ & $0.66^{+0.09}_{-0.07}$ &
$0.59^{+0.13}_{-0.06}$
\\
\noalign{\smallskip}
\noalign{\smallskip}
A$_{\rm N-D}$ ($^7$Be) (\%)
& ---
& $0.0^{+0.1}_{-0.0}$
& $ 15^{+17}_{-15}$\\
\noalign{\smallskip} \noalign{\smallskip} $[$R ($^8$B)$]$ $\nu-e$
scattering
& & &\\
($T_{\rm th}=3.5$ MeV) & $0.46\pm 0.03$ & $0.46\pm 0.04$
& $0.45\pm 0.03$ \\
($T_{\rm th}=5$ MeV) & $0.46\pm 0.03$ & $0.46\pm 0.04$
& $0.45\pm 0.03$ \\
$[$CC$]$ (KamLAND)
& & &\\
($E_{\rm th}=2.72$ MeV)
&$0.56^{+0.14}_{-0.22}$
&$0.56^{+0.20}_{-0.34}$
& ---
\\
($E_{\rm th}=1.22$ MeV)
&$0.57^{+0.1}_{-0.18}$
&$0.57^{+0.16}_{-0.31}$
& ---
\\
\noalign{\smallskip}
\noalign{\smallskip}
$\delta E_{\rm visible}$ (KamLAND) (\%)
& & &\\
($E_{\rm th}=2.72$ MeV)
& $-7^{+12}_{-2}$
& $-7^{+14}_{-4}$
& ---\\
($E_{\rm th}=1.22$ MeV)
& $-7^{+11}_{-4}$
& $-7^{+15}_{-7}$
& ---\\
\noalign{\smallskip}
\noalign{\smallskip}
$\delta\sigma$ (KamLAND) (\%)
& & &\\
($E_{\rm th}=2.72$ MeV)
& $-6^{+16}_{-9}$
& $-6^{+21}_{-12}$
& ---\\
($E_{\rm th}=1.22$ MeV)
& $-9^{+19}_{-7}$
& $-9^{+28}_{-11}$
& ---\\
\noalign{\smallskip} \noalign{\smallskip} $[p$-$p]$
 $\nu-e$ scattering & & &\\
($T_{\rm th}=100$ keV) & $0.722^{+0.033}_{-0.042}$
& $0.722^{+0.085}_{-0.067}$
& $0.689^{+0.058}_{-0.065}$ \\
($T_{\rm th}=50$ keV) & $0.718^{+0.034}_{-0.043}$
& $0.718^{+0.086}_{-0.069}$
& $0.687^{+0.058}_{-0.068}$ \\
\noalign{\smallskip} \hline
\end{tabular}
}

Based upon the results given in table~\ref{tab:afterpredictions},
we predict that SNO will not measure a statistically
significant ($> 3\sigma$) distortion to the recoil energy spectrum
for the CC reaction. This prediction constitutes an important
consistency test of the oscillation analysis and the understanding
of systematic effects in the detector.

The prediction for the reduced $^7$Be $\nu-e$ scattering rate,
\begin{equation}
[{\rm ^7Be}] ~\equiv~ \frac{\rm Observed ~\nu-e ~scattering~
rate}{\rm BP00~ predicted ~rate}~,  \label{eq:be7reducedrate}
\end{equation}
is remarkably precise and remarkably stable. The current
prediction for what BOREXINO will measure if the LMA solution is
valid  is $[{\rm ^7Be}] = 0.64^{+0.04}_{-0.03}$. A somewhat
smaller value is predicted if the LOW solution is correct.

We have also evaluated the predictions for a generic $p-p$
neutrino-electron scattering detector. The reduced rate for this
detector is defined, analogously to the reduced rate for $^7$Be
detectors, by the relation
\begin{equation}
[p-p] ~\equiv~ \frac{\rm Observed ~\nu-e ~scattering~ rate}{\rm
BP00~ predicted ~rate}~.  \label{eq:ppreducedrate}
\end{equation}
We present the predicted rate for two plausible kinetic energy
thresholds, $100$ keV and $50$ keV. The predicted rate is precise
and robust, which supports previous suggestions that a measurement
of the $p-p$ neutrino scattering rate can be used to determine
accurately the dominant solar neutrino mixing angle.

We also include in table~\ref{tab:afterpredictions} the
predictions for the expected  $\nu-e$ scattering rate, [R($^8$B)]
(defined similarly to [$^7$Be] and [$p-p$]), from $^8B$ neutrinos
above two different electron recoil kimetic-energy energy
thresholds, $T_{\rm th}=3.5$ MeV and $T_{\rm th}=5$
MeV\footnote{If we normalize the detector exposure to have $55$
events a day in the energy window $0.2 {\rm ~MeV} < T < 0.8 {\rm
~MeV}$ (G. Bellini, private communication, 5/2002) and assume that
the detector efficiency is energy independent (but include of
course the energy dependence of the neutrino cross section and the
detector resolution of BOREXINO), we estimate for the BP00 $\nu_e$
flux a total of $174$ events a year above $3.5$ MeV of which $109$
would be above $5$ MeV.}. The rates expected if the LMA or LOW
solution is correct are essentially identical to the
neutrino-electron scattering rates measured at higher thresholds
by the SNO and Super-Kamiokande collaborations, which are $0.47
\pm 0.13$ and $0.46 \pm 0.04$, respectively (relative to the value
predicted with the BP00 standard solar model flux). Thus a
measurement of the $^8$B $\nu-e$ scattering rate in BOREXINO, or
with a low threshold in Super-Kamiokande, is not expected to yield
a significant deviation from the rate measured at higher energies.

The predicted value of the reduced CC event rate in the KamLAND
reactor experiment,
\begin{equation}
[{\rm CC}] ({\rm KamLAND}) ~\equiv~ \frac{\rm Observed ~\bar{\nu} +
p ~absorption~ rate}{\rm No ~ oscillation ~rate}~,
\label{eq:kamlandreducedrate}
\end{equation}
is not significantly affected by the recent SNO results. The
best-fit prediction shifts slightly (to a lower value) but the
shift is well-within the $1\sigma$ currently allowed range.

For KamLAND, it is convenient to
 represent the distortion of the visible energy spectrum by
the fractional deviation from the undistorted spectrum of the
first two moments of the energy spectrum. We follow the notation
and analysis of refs.~\cite{robust,tencommand,bkl97}.  The
predicted fractional distortion of the first two moments is not
significantly affected by the recent SNO measurements. Since
table~\ref{tab:afterpredictions}  and
table~\ref{tab:beforepredictions} give only the fractional changes
of the moments relative to the moments for the undistorted
spectrum, we must also specify the values calculated for a
spectrum unaffected by new physics. In the absence of
oscillations, one expects~\cite{robust}: $\langle E_{\rm
vis}\rangle_0=3.97$ MeV and $\langle\sigma\rangle_0=1.26$ MeV for
$E_{\rm threshold} = 1.22$ MeV ($\langle E_{vis}\rangle_0=4.33$
MeV and $\langle\sigma\rangle_0=1.06$ MeV for $E_{\rm threshold}
=1.72$ MeV).

\subsection{Predictions for vacuum solutions}
\label{subsec:predictionsvac}

The clearest evidence against the vacuum oscillations would be the
observation of a rate depletion or a spectral distortion in the
KamLAND reactor experiment. The $\Delta m^2$ for vacuum
oscillations is too small to lead to an observable effect with
KamLAND.

 Within the
$3\sigma$ allowed VAC regions, the distortion of the SNO spectrum
corresponds to a shift in the first (second) moment of the recoil
energy distribution of at most -2\% (+6\%) and no significant
seasonal variation at SNO is expected. The predicted $^7$Be  rate
for vacuum oscillations at BOREXINO is in the same range as the
predictions for LMA and LOW solutions. The most striking signal
for vacuum oscillations would be the observation of a large
seasonal variation in the BOREXINO experiment, with a clear
pattern of the monthly dependence of the observed rate
\cite{borexseasonal}. There should also be a day-night effect at
BOREXINO associated with this seasonal variation; the day-night
asymmetry should be  at most $\pm 8$\% (the size and
sign of this asymmetry is sensitive to the exact value of $\Delta
m^2$ considered to be within the allowed VAC islands) due to the
dependence of the survival probability upon the earth-sun
distance.

\FIGURE[!t]{ \centerline{\psfig{figure=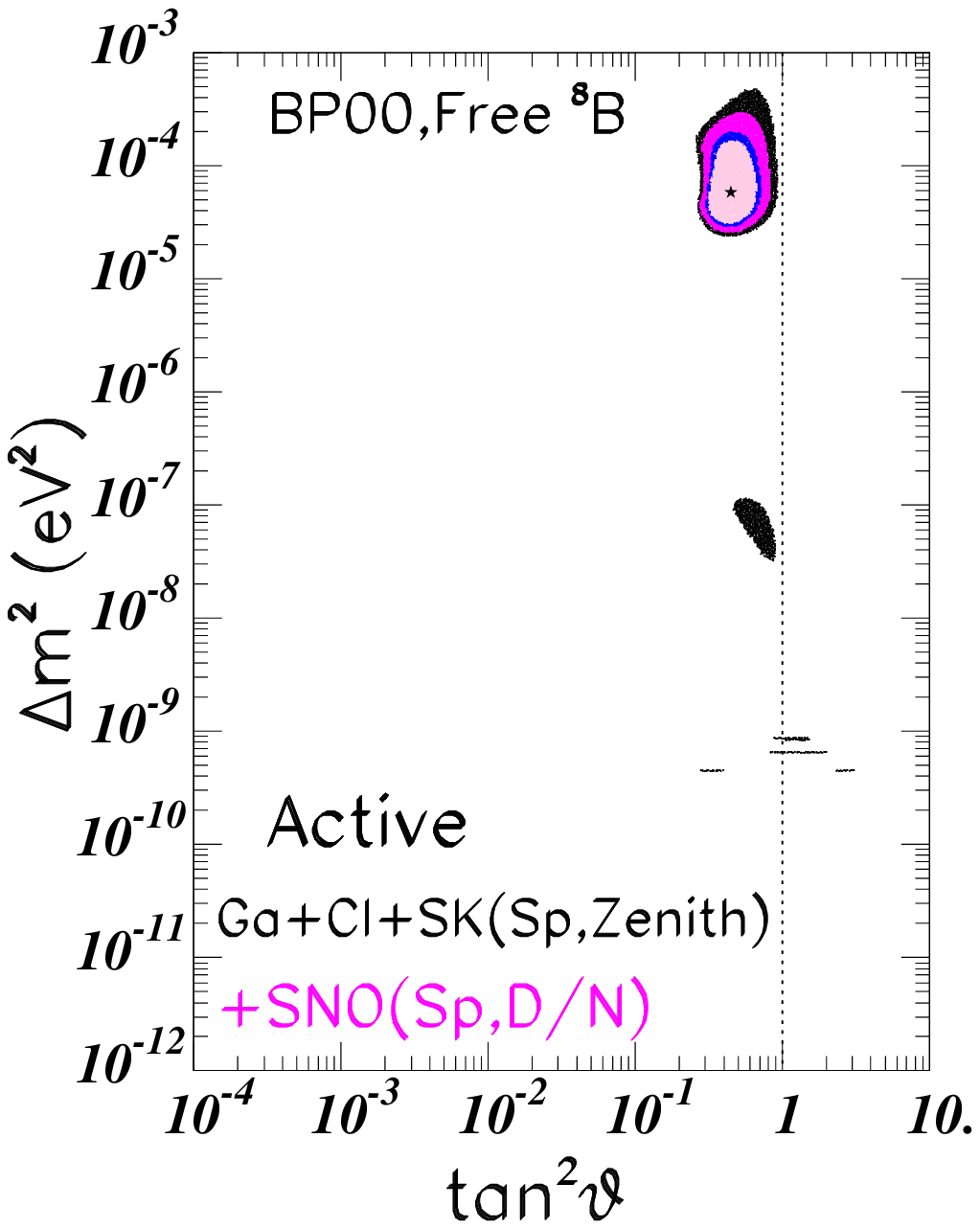,width=3in}}
\caption{{\bf Global neutrino oscillation solutions with full SNO
day-night spectrum and analysis strategy (a).} To assess the
robustness of the global oscillation solutions, compare this
figure with figure~\ref{fig:global3}a or \ref{fig:beforeafter}b.
The input data used in constructing figure~\ref{fig:globalnew}
include the SNO day-night spectrum ~\cite{snodaynight}, the
Chlorine~\cite{chlorine} and Gallium
(averaged)~\cite{sage2002,gallex,gno,sage} event rates, as well as
the zenith angle-recoil energy spectrum data presented by
Super-Kamiokande~\cite{superk} and the neutrino fluxes and
uncertainties predicted by the standard solar model~\cite{bp2000}
except for the $^8$B flux (which is treated as a free parameter).
The rates from the GALLEX/GNO and SAGE experiments have been
averaged to provide a unique data point ($72.4 \pm 4.7$ SNU). The
CL contours shown in the figure are $90$\%, $95$\%, $99$\%, and
$99.73$\% ($3\sigma$). The global best-fit points are marked by a
star. \label{fig:globalnew}} }

\section{Global analysis including the full SNO day-night energy spectrum}
\label{sec:fullsno}

In order to help determine the robustness of the oscillation
solutions found in section~\ref{sec:global}, we have recalculated
the global solutions using the full SNO day-night energy spectrum
instead of the two step treatment of SNO data described in
section~\ref{subsec:aficionados}. This full-SNO procedure has been
used, by among others, the SNO collaboration~\cite{snodaynight}.

We begin by presenting in section~\ref{subsec:globalfullsno} the
global solutions obtained using the full SNO analysis procedure
and then compare in section~\ref{subsec:comparison} the allowed
ranges for $\Delta m^2$, $\tan^2 \theta$, and $f_{\rm B}$ found by
the two step procedure and the inclusion of the full SNO day-night
energy spectrum. We compare in
section~\ref{subsec:predictionsfullsno} the predictions for future
measurements with the BOREXINO, KamLAND, and SNO detectors and a
generic future $p-p$ solar neutrino detector. We describe in
section~\ref{subsec:aficionadosFF} details of the analysis
procedure, details that are probably only of interest to
aficionados.

\subsection{Global solutions}
\label{subsec:globalfullsno}

 Figure~\ref{fig:globalnew} presents
the result of our "full-SNO" global analysis (strategy a) that was
obtained using the full SNO day-night energy spectrum in
combination with the Super-Kamiokande zenith-angle energy spectrum
and the Ga and Cl event rates. The number of degrees of freedom in
this analysis is 77: 44 (Super-Kamiokande zenith-angle energy
spectrum) + 2 (Ga and Cl rates) + 34 (SNO day-night energy
spectrum) $-$3 parameters ( $\Delta {\rm m}^2$, $\theta$, and
$f_{\rm B}$).

To simplify the comparison with our results obtained with the
two-step analysis of the SNO data (see figure~\ref{fig:global3}
and figure~\ref{fig:beforeafter}b),  we continue to use the
average gallium rate $72.4\pm4.7$ SNU and do not include the
recent, preliminary GNO data reported at Neutrino 2002 (which
would result into a slight lowering of the rate to $70.8\pm4.4$
SNU). We have verified that the use of this lower rate would not
affect significantly any of the conclusions of the present
analysis. The $\chi^2$ value for LMA would be lowered by $+0.3$,
while the difference between LMA and LOW (and VAC) solutions is
decreased by $0.3$.
\TABLE[h!!t]{\caption{\label{tab:bestfitsnew} {\bf Best-fit global
oscillation parameters obtained using the full SNO day-night
energy spectrum.} The values given here, which were computed using
the full SNO day-night energy spectrum, should be compared with
the values given in table~\ref{tab:bestfitsa}, which were computed
using the two-step treatment of SNO data that is described in
section~\ref{subsec:aficionados}. Table~\ref{tab:bestfitsnew}
gives the best-fit values for $\Delta m^2$, $\tan^2 \theta$,
$\chi^2_{\rm min}$, and g.o.f. for all the oscillation solutions
among active solar neutrinos that have been previously discussed
(see, e.g., ref.~\cite{bgp}). The quantity $f_{\rm B}$ measures
the $^8$B solar neutrino flux in units of the predicted BP00
neutrino flux, see eq.~(\ref{eq:fbdefn}). The oscillation
solutions are obtained by varying the $^8$B flux as free parameter
in a consistent way: simultaneously in the rates and in the fits
to the night and day energy spectra. The differences of the
squared masses are given in ${\rm eV^2}$. The number of degrees of
freedom is is 77 [44 (Super-Kamiokande zenith-angle energy
spectrum) + 2 (Ga and Cl event rates) + 34 (SNO day-night energy
spectrum) $-$3 (parameters: $\Delta {\rm m}^2$, $\theta$, and
$f_{\rm B}$)]. The goodness-of-fit given in the last column is
calculated relative to the minimum for each solution. (Solutions
that have $\chi_{\rm min}^2 \geq 75.4 + 11.8=87.2$ are not allowed
at the $3\sigma$ CL) }
\begin{tabular}{lccccc}
\noalign{\bigskip} \hline \noalign{\smallskip} Solution&$\Delta
m^2$&$\tan^2 (\theta)$& $f_{\rm B,best}$
& $\chi^2_{\rm min}$ & g.o.f. \\
\noalign{\smallskip} \hline \noalign{\smallskip} LMA&
$5.8\times10^{-5}$  &$4.5\times10^{-1}$& 1.02
& 75.4 &$53$\% \\
LOW& $7.9\times10^{-8}$  &$6.3\times10^{-1}$& 0.93
& 85.0 &$25$\%\\
VAC& $6.5\times10^{-10}$  &$1.5\times10^{0}$& 0.76
& 85.5 &$23$\%\\
SMA& $4.7\times10^{-6}$  &$1.7\times10^{-3}$& 0.92
& 101.0 &$3.4$\%\\
Just So$^2$ & $5.8\times10^{-12}$  & $1.0\times10^{0} $& 0.45
& 115.1 &$0.3$\%\\
Sterile VAC & $4.5\times10^{-10}$ & $2.5\times10^{0}$ & 0.78
& 101.4 &$3.3$\%\\
Sterile Just So$^2$ & $5.8\times10^{-12}$  &\hskip
-6pt$1.0\times10^{0} $ & 0.45
& 111.8 &$0.6$\%\\
Sterile SMA & $3.6\times10^{-6}$ & $4.2\times10^{-4}$ & 0.55
& 115.1 &$0.3$\%\\
\noalign{\smallskip} \hline
\end{tabular}}

\subsection{Comparison of two-step and full-SNO analysis
procedures} \label{subsec:comparison}

 Table~\ref{tab:bestfitsnew} gives, for our full-SNO
global analysis (cf. figure~\ref{fig:globalnew}), the best-fit
values for $\Delta m^2$ and $\tan^2 \theta$ for all the neutrino
oscillation solutions. The table also lists the values of
$\chi_{\rm min}^2$ for each solution. The regions for which the
local value of $\chi_{\rm min}^2$ exceeds the global minimum  by
more than $11.83$ are not allowed at $3\sigma$ CL.

Comparing the right hand panel of figure~\ref{fig:beforeafter} with
figure~\ref{fig:globalnew}  and the corresponding minima in table
~\ref{tab:bestfitsa} with those in  table~\ref{tab:bestfitsnew}
shows that the main effect of the inclusion of the full day-night
spectrum information is, as expected, the  moderate worsening of
solutions with larger spectrum distortions. SMA becomes even
further disfavored (acceptable only at the $4.7\sigma$ CL) while
vacuum solutions are now only marginally allowed at 3$\sigma$
(acceptable at the $2.7\sigma$ CL). Furthermore, the best-fit
vacuum solution is moved into the second lowest $\Delta m^2$
island. The confidence level for the LOW solution exceeds $99$\% CL
(acceptable at the $2.6\sigma$ CL).

How are the ranges of allowed masses and mixing changed?

In units of ${\rm eV^2}$, we find for the LMA solution the
following $3\sigma$ limits on $\Delta m^2$,
\begin{equation}
2.4\times 10^{-5} < \Delta m^2 < 4.7 \times 10^{-4} .
\label{eq:newmasslimitlma}
\end{equation}
For the LOW solution, the allowed mass range  is
\begin{equation}
3.2\times 10^{-8} < \Delta m^2 < 1.1 \times 10^{-7}.
\label{eq:newmasslimitlow}
\end{equation}
Comparing these results with the corresponding results given in
eq.~\ref{eq:masslimitlma} and eq.~\ref{eq:masslimitlow}  for the
two-step procedure for analyzing SNO data, we see that in both
cases the full-SNO procedure gives a slightly larger $3\sigma$
range for $\Delta m^2$.

The allowed range for the LMA mixing angle is
\begin{equation}
 0.27 < \tan^2 \theta < 0.92
 \label{eq:newtanlimitlma}
 \end{equation}
and for the LOW solution
\begin{equation}
 0.47< \tan^2 \theta < 0.89
\label{eq:newtanlimitlow}
\end{equation}
Comparing the limits given above  with the values found earlier
(see eq.~\ref{eq:tanlimitlma} and eq.~\ref{eq:tanlimitlow}) using
the two-step procedure, we conclude that the full-SNO procedure
gives a slightly shifted $3\sigma$ range for $\tan^2 \theta$.

For the global solution shown in figure~\ref{fig:globalnew}, the
$1\sigma[3\sigma]$ allowed range of $f_{\rm B}$ is
\begin{equation}
 f_{\rm B} ~=~ 1.02\pm 0.08~({1\sigma})~~[f_{\rm B} ~=~
 1.02\pm{0.24}]~({3\sigma})~({\rm LMA}).
\label{eq:fbrangelmanew}
\end{equation}
 The range of $f_{\rm B}$ was calculated,  as just as it was for
eq~.(\ref{eq:fbrangelma}) and eq.~(\ref{eq:fbrangelow}), by
marginalizing over the full space of oscillation parameters
($\Delta m^2$, $\tan^2 2\theta$). Hence we use $\Delta\chi^2$
conditions for 1 dof, relative to the global minimum value for
$\chi^2$ that lies in the LMA allowed region. Since for the
full-SNO analysis procedure $\Delta \chi^2 = \chi^2_{\rm LOW}
-\chi^2_{\rm LMA}>9$, there are no allowed solutions for $f_{\rm
B}$ at the $3\sigma$ CL ($1$\,dof) within either the LOW or vacuum
solution domains.

The uncertainties shown in eq.~\ref{eq:fbrangelmanew} are
essentially the same as the uncertainties in $f_{\rm B} $ that
were found using the two-step procedure for analyzing SNO data
(see eq~.\ref{eq:fbrangelma} and eq.~\ref{eq:fbrangelow}). We
conclude that the uncertainties in all three of the parameters
$\Delta m^2$, $\tan^2 \theta$, and $f_{\rm B}$ are robust with
respect to the choice of analysis procedures for the SNO data.
\subsection{Predictions for BOREXINO, KamLAND, SNO and a generic $p-p$ detector}
\label{subsec:predictionsfullsno}
The predicted event rates for measurable solar neutrino quantities
are essentially the same if one uses the full-SNO day-day night
energy spectrum or if one uses the two-step procedure.

Table~\ref{tab:afterpredictionsnew} gives the predicted ranges for
future solar neutrino observables that correspond to the allowed
regions in figure~\ref{fig:globalnew}. The values shown in
table~\ref{tab:afterpredictionsnew} should be compared with the
results presented in table~\ref{tab:afterpredictions}, which was
obtained using the two-step procedure for SNO data. The comparison
shows that there are small shifts in some of the central values of
the observables due to the slight shifts in the positions of the
LMA and LOW best fit points. However, the predicted ranges are not
significantly modified.

\TABLE[!t]{\caption{\label{tab:afterpredictionsnew} {\bf ``After''
Predictions obtained using the full SNO day-night energy
spectrum.} This table presents for future solar neutrino
observables the best-fit predictions and $1\sigma$ and $3\sigma$
ranges that were obtained by using the analysis including the SNO
day-night spectrum and all other solar neutrino data currently
available. The best-fit values and uncertainties given here
correspond to the allowed regions in the
figure~\ref{fig:globalnew}. The predictions given in this table
should be compared with table~\ref{tab:afterpredictions}, which
was obtained by treating the SNO data by the two-step procedure
described in section~\ref{subsec:aficionados}.}
\begin{tabular}{lccc}
\noalign{\bigskip} \hline \noalign{\smallskip} Observable&  b.f.
$\pm 1\sigma$  &LMA  $\pm 3\sigma$  & LOW  $\pm 3\sigma$
\\
\noalign{\smallskip} \hline \noalign{\smallskip} A$_{\rm N-D}$
(SNO CC) (\%) & $4.2^{+3.8}_{-2.9}$ & $4.2^{+8.9}_{-4.2}$ &
$2.7^{+2.2}_{-2.2}$
\\
\noalign{\smallskip}\noalign{\smallskip}$\delta T$ (SNO CC) (\%) &
$-0.10^{+0.30}_{-0.50}$ & $-0.10^{+0.38}_{-1.64}$&
$0.46^{+0.52}_{-0.41}$
\\
\noalign{\smallskip}\noalign{\smallskip}$\delta\sigma$ (SNO CC)
(\%)& $-0.25^{+0.28}_{-0.50}$ & $0.25^{+0.39}_{-1.46}$ &
$0.38^{+0.42}_{-0.33}$
\\
\noalign{\smallskip} \noalign{\smallskip} $[$R ($^7$Be)$]$ $\nu-e$
scattering & $0.64\pm 0.03$ & $0.64^{+0.09}_{-0.05}$ & $0.59\pm
0.04$
\\
\noalign{\smallskip} \noalign{\smallskip} A$_{\rm N-D}$ ($^7$Be)
(\%) & $0.0^{+0.0}_{-0.0}$ & $0.0^{+0.1}_{-0.0}$
& $ 23^{+9}_{-13}$\\
\noalign{\smallskip} \noalign{\smallskip}$[$R ($^8$B)$]$ $\nu-e$
scattering
& & &\\
($T_{\rm th}=3.5$ MeV) & $0.46\pm 0.03$ & $0.46\pm 0.04$
& $0.45\pm 0.03$ \\
($T_{\rm th}=5$ MeV) & $0.45\pm 0.03$ & $0.45\pm 0.04$
& $0.45\pm 0.03$ \\
\noalign{\smallskip} \noalign{\smallskip} $[$CC$]$ (KamLAND)
& & &\\
 ($E_{\rm th}=2.72$ MeV) &$0.58^{+0.10}_{-0.27}$
&$0.58^{+0.14}_{-0.35}$ & ---
\\
($E_{\rm th}=1.22$ MeV) &$0.57^{+0.08}_{-0.20}$
&$0.57^{+0.12}_{-0.30}$ &
---
\\
\noalign{\smallskip} \noalign{\smallskip} $\delta E_{\rm visible}$
(KamLAND) (\%)
& & &\\
($E_{\rm th}=2.72$ MeV) & $-7^{+13}_{-2}$ & $-7^{+14}_{-4}$
& ---\\
($E_{\rm th}=1.22$ MeV) & $-5^{+10}_{-5}$ & $-5^{+13}_{-8}$
& ---\\
\noalign{\smallskip} \noalign{\smallskip} $\delta\sigma$ (KamLAND)
(\%)
& & &\\
($E_{\rm th}=2.72$ MeV) & $-11^{+23}_{-4}$ & $-11^{+26}_{-7}$
& ---\\
($E_{\rm th}=1.22$ MeV) & $-14^{+24}_{-3}$ & $-14^{+32}_{-6}$
& ---\\
\noalign{\smallskip} \noalign{\smallskip} $[p$-$p]$ $\nu-e$
scattering
& & &\\
($T_{\rm th}=100$ keV) & $0.699^{+0.026}_{-0.025}$ &
$0.699^{+0.069}_{-0.044}$
& $0.686^{+0.033}_{-0.034}$ \\
($T_{\rm th}=50$ keV) & $0.695^{+0.026}_{-0.025}$ &
$0.695^{+0.070}_{-0.045}$
& $0.681^{+0.035}_{-0.037}$ \\
\noalign{\smallskip} \hline
\end{tabular}}

\subsection{Analysis details for aficionados: using the full SNO day-night energy spectrum}
\label{subsec:aficionadosFF} We describe in this subsection the
details of our analysis procedure when we include the full SNO
day-night energy spectrum. We perform a direct fit to the total
SNO day-night energy spectrum, including contributions from  NC,
CC, and ES events.

In order to avoid redundancy with the discussion in
section~\ref{subsec:aficionados}, we also provide in this section
some detailed information about analysis procedures and error
estimates that is necessary for comparing our results with the
results obtained by other analysis groups. This additional
information is relevant to both the two-step and the full-SNO
analysis procedures.

We use the detailed information provided in ref.~\cite{snourl} by
the SNO collaboration. For each point in oscillation parameter
space, we compute (as with the two-step procedure) the expected
event rates for the chlorine and gallium experiments, and the
Super-Kamiokande zenith-angle energy spectrum. For the SNO
day-night energy spectrum, we compute the NC, CC, and ES
contributions in each energy bin.  We construct a theoretical SNO
spectrum by summing the NC, CC, and ES contributions together with
the background rates given in ref.~\cite{snourl}. With these model
predictions, we make a global fit including as the $80$ data
points, the $2$ radio-chemical rates, the $44$ data points of the
Super-Kamiokande zenith-angle spectrum,  and the $34$ points of
the  SNO day-night spectrum.

We include the errors and their correlations for all of these
observables. We construct an $80 \times 80$ covariance error
matrix that includes the effect of correlations between the
different errors as off diagonal elements (an alternative ``pull''
method has been recently proposed in ref.~\cite{foglinc}). The
main sources of correlations are due to the BP00 neutrino fluxes,
the cross section uncertainties for the gallium and chlorine
rates, and the cross sections and experimental systematic errors
for Super-Kamiokande and SNO energy spectra. As in our two-step
analysis, the flux and cross section errors are included following
the general approach described in ref.~\cite{fogli} with the
refinements described in the appendix of ref.~\cite{sterile}.

The experimental systematic errors can be either correlated or
uncorrelated in energy. We  include these errors as follows
\footnote{ We describe here our error treatment for both
Super-Kamiokande and SNO energy spectra. For Super-Kamiokande, the
treatment is the same for the two-step analysis and for the
full-SNO procedure.}. For Super-Kamiokande,  we take from
table~$1$ of ref.~\cite{smy2002} the systematic error, which we
include as uncorrelated among the Super-Kamiokande energy bins but
fully correlated among the zenith-angle bins. For SNO, energy
independent but correlated errors include the systematic errors
from the uncertainty in vertex reconstruction ($+3$\% for CC and
ES and +$1.45$\% for NC),  the background errors from neutron
capture, and low-energy correlated systematic uncertainties. Each
of these errors are fully correlated among the $34$ SNO bins.
Furthermore, the fractional error is assumed to be constant, i.e.
independent (in \%) of the oscillation point and affecting equally
$^8$B and hep neutrinos.

For both Super-Kamiokande and SNO there also energy-dependent and
energy-correlated errors from (i) the $^8$B spectral energy shape,
(ii) the absolute energy scale,  and (iii) the energy resolution.
The fractional value of those errors is, in general, affected by
distortions of the spectrum due to oscillations and therefore they
should  be evaluated for each point in the oscillation parameter
space. The Super-Kamiokande and the SNO collaborations do not
specify the details of what they use for these errors. We evaluate
the three errors listed above as follows.  First, we compute for
each point in oscillation parameter space the expected number of
ES events in each energy bin for the Super-Kamiokande spectrum,
and the number of CC, ES, and NC events in each energy bin for
SNO. We then evaluate the fractional change in the calculated
results by: (i) shifting the normalization of the $^8$B spectrum
up and down within the 1$\sigma$ error determined in
ref.~\cite{chlorineb8cs} in order to obtain the error associated
with the  $^8$B spectrum shape; (ii) shifting the calibrated
energy $T'$ to $T'(1\pm\delta_T)$ with $\delta T= 0.64$\%
(1.21\%)  for Super-Kamiokande (SNO) to obtain the error
associated with the energy scale; (iii) sifting the width of the
resolution function $\sigma$ to $\sigma(1\pm\delta_\sigma)$ with
$\delta_\sigma = 0.025$ for Super-Kamiokande($\delta_\sigma =
0.045+0.004\times (T-4.98)$ for SNO) to obtain the error due to
the energy resolution. Whenever a comparison was possible, we have
verified that the error estimates made as described above are in
good agreement with the information about the errors that has been
provided by the Super-Kamiokande and SNO collaborations. Each of
these errors is fully correlated among the different bins in each
experiment. Moreover, the shape errors are also correlated between
the Super-Kamiokande and SNO experiments.

\section{Discussion and summary}
\label{sec:discuss}

In the following, we present without parentheses results that were
obtained by the two-step procedure for analyzing SNO data (and
present in parentheses the corresponding results that were
obtained by analyzing the full SNO day-night energy spectrum). If
the results are identical for both treatments of the SNO data,
then we omit the value in parenthesis. By comparing the values
given with and without parentheses, the reader can see the rather
small differences that result from different choices of how to
treat the SNO data.

Figure~\ref{fig:global3} and figure~\ref{fig:beforeafter},
together with table~\ref{tab:bestfitsa} (or, alternatively,
figure~\ref{fig:globalnew} and table~\ref{tab:bestfitsnew}), tell
much of the story. These global analyses, which use all the
available solar neutrino data, demonstrate that only large mixing
angle solutions are currently allowed at the $3.7\sigma$
($4.7\sigma$) confidence level.
We therefore know at an impressive confidence level that there are
relatively large mixing angles for oscillations between solar
neutrinos (and between atmospheric neutrinos), unlike the small
mixing angles among quarks. However, precise maximal mixing is
excluded at $3.2\sigma$ for MSW solutions [see
eq.~\ref{eq:tanlimitlma} and eq.~\ref{eq:tanlimitlow}
(eq.~\ref{eq:newtanlimitlma} and eq.~\ref{eq:newtanlimitlow}) for
the exact allowed regions] and at $2.8\sigma$ ($2.9\sigma$) for
vacuum oscillations. The  masses and mixing angles for solar
neutrinos do not appear to satisfy an obviously simple pattern.

Among the MSW solutions, the situation has also been clarified.
The LMA solution is now the only viable solution at a level of
$2.5\sigma$ (2.6 $\sigma$). The LOW solution is excluded at the
{\bf $98.8$\%}  ($99.2$\%) CL. The KamLAND and BOREXINO
experiments will test strongly this conclusion. The SMA solution
is now excluded at more than $3.7\sigma$  (4.7 $\sigma$). Pure
sterile oscillations are excluded at $5.4\sigma$ (4.7 $\sigma$).

The possibility of a strictly  energy independent solution of the
solar neutrino problem is excluded at more than $3.6\sigma$. If we
consider solutions that have an energy variation less than $10$\%
over the energy range of interest, the exclusion is at more than
$2.6\sigma$. We note that such a weak energy dependence  can
probably not be distinguished experimentally  from strict energy
independence  (see ref.~\cite{gp} for details).

Vacuum solutions are not favored, but are acceptable at
$2.1\sigma$ ($2.7\sigma$)\footnote{We have attempted to determine
why the SNO collaboration did not find any $3\sigma$ allowed
vacuum solutions. We have found at least one plausible
explanation. If we make the imprecise approximations that are
listed below,  then we also do not find any  allowed vacuum
solutions at the $3\sigma$ CL. Here are the approximations (with
references to descriptions of more accurate error treatments) : 1)
neglect the uncertainty~\cite{chlorineb8cs} in the true shape of
the  $^8$B undistorted energy spectrum, 2) neglect the dependence
on the neutrino oscillation parameters of the calculated errors
for the SNO and Super-Kamiokande  spectral energy distributions
(i. e., compute the errors for an undistorted $^8$B energy
spectrum so that the percentage errors for each energy bin are
independent of $\Delta m^2$ and $\tan^2 \theta$ ), and 3) neglect
the correlations and energy dependence of the
errors~\cite{sterile, GaCS} for the chlorine and gallium neutrino
cross sections. Making these approximations, we find that the
$\Delta \chi^2$ between the best-fit LMA solution and the vacuum
solution with the lowest $\chi^2$ is $14.6$ for the two-step SNO
procedure and is $12.3$ for the analysis including the full SNO
day-night energy spectrum. Both of these values exceed the $\Delta
\chi^2$ of $11.83$ that corresponds to a $3\sigma$ CL. We also
tested the effect of neglecting the energy shape~\cite{be7shape}
of the $^7$Be line. This approximation turned out to be important
for predicting the values of future observables for BOREXINO and
KamLAND, but does not significantly affect the computed difference
between the global best-fit values of $\chi^2$ between LMA and
vacuum solutions.}.

\TABLE[h!!t] { \centering \caption{{\bf Neutrino oscillation
predictions for the chlorine and gallium radiochemical
experiments.}  The predictions are based upon the global analysis
strategy (a), and use the neutrino fluxes for the BP00 standard
solar model (except for the $^8$B flux for which the factor
$f_{\rm B}$ is included) and  the neutrino absorption cross
sections~\cite{neutrinoastrophysics,chlorineb8cs,GaCS}. The rates
are presented for the best-fit oscillation parameters of the
allowed solutions listed in table~\ref{tab:bestfitsa}. The total
rates should be compared with the standard solar model
values~\cite{bp2000}, which are, $7.6^{+1.3}_{-1.1}$ (chlorine)
and $128^{+9}_{-7}$(gallium), and the measured values, which are
$2.56\pm 0.23$ (chlorine~\cite{chlorine}) and $72.4\pm
4.7$(gallium~\cite{sage2002,gallex,gno}.
)\label{tab:radiochemical} }
\begin{tabular}{@{\extracolsep{-5pt}}lcccccc}
\hline \noalign{\smallskip}
Source&Cl&Ga&Cl&Ga&Cl&Ga\\
&(SNU)&(SNU)&(SNU)&(SNU)&(SNU)&(SNU)\\
&LMA&LMA&LOW&LOW&VAC&VAC\\
\noalign{\smallskip} \hline \noalign{\smallskip}
pp            &  0   & 40.4 &  0   & 38.2 &  0   & 40.3\\
pep           & 0.12 & 1.51 & 0.10 & 1.25 & 0.15 & 1.82\\
hep           & 0.01 & 0.02 & 0.02 & 0.03 & 0.02 & 0.04\\
${\rm ^7Be}$  & 0.62 & 18.6 & 0.53 & 16.0 & 0.46 & 14.0\\
${\rm ^8B}$   & 2.05 & 4.35 & 2.26 & 4.72 & 2.34 & 5.00\\
${\rm ^{13}N}$& 0.04 & 1.79 & 0.04 & 1.56 & 0.05 & 1.83\\
${\rm ^{15}O}$& 0.15 & 2.83 & 0.15 & 2.44 & 0.18 & 3.01\\
${\rm ^{17}F}$& 0.00 & 0.03 & 0.00 & 0.03 & 0.00 & 0.04\\
\noalign{\medskip}
&\hrulefill&\hrulefill&\hrulefill&\hrulefill&\hrulefill
&\hrulefill
\\
Total         & $3.03 $ & $69.6 $ & $3.11$&
$64.2$& $3.21$ & $66.0$\\
\noalign{\smallskip} \hline
\end{tabular} }

For the reader who (like us) often prefers pictures to tables,
figure~\ref{fig:pee} shows the predicted energy dependence and
day-night difference for the current best-fit LMA, LOW, and VAC
solutions.

The global oscillation solutions constrain the active $^8$B solar
neutrino flux with somewhat greater accuracy than the SNO NC
measurement.  The accuracy of the SNO NC measurement is $\pm
12$\%~\cite{snonc}. We find from a global oscillation solution of
all the data, that at the $1\sigma$ CL, $f_{\rm B} ~=~ 1.07 \pm
0.08 $ ($f_{\rm B} ~=~ 1.02 \pm 0.08$), where $f_{\rm B}$ is the
total flux of active $^8$B solar neutrinos in units of the flux
predicted by the standard solar model~\cite{bp2000}. The allowed
ranges of $f_{\rm B}$ for the LMA and LOW solutions are given in
eq.~(\ref{eq:fbrangelma})  and eq.~(\ref{eq:fbrangelow})
(eq.~\ref{eq:fbrangelmanew}) at $1\sigma$ and $3\sigma$ CL.

We have calculated the predicted best-fit values and allowed
ranges for ten important solar neutrino observables for the
BOREXINO, KamLAND, and SNO experiments and for a generic detector
of $\nu-e$ scattering by $p-p$ neutrinos. The observables include
total rates, day-night effects, and spectral distortions. The
predictions are summarized in the ``Before'' and ``After'' tables,
table~\ref{tab:afterpredictions} and
table~\ref{tab:beforepredictions}, and in
section~\ref{subsec:predictionsvac}. Similar results are presented
for the full-SNO procedure in
table~(\ref{tab:afterpredictionsnew}) and in
section~\ref{subsec:predictionsfullsno}.
 The predictions are rather robust and are not significantly affected
  by the choice of procedure (two-step or full-SNO)
 used to analyze the SNO data. Only the
predictions for the disfavored LOW solution are significantly
affected by the recent SNO measurements~\cite{snonc,snodaynight}.
We have also calculated the predicted values and ranges given in
table~\ref{tab:afterpredictions} and
table~\ref{tab:beforepredictions} both with and without taking
account of potential spectral energy distortions in interpreting
the SNO data (see section~\ref{subsec:aficionados} and
ref.~\cite{snourl}). Both methods give essentially identical
results.

For completeness, table~\ref{tab:radiochemical} gives the
individual contributions of different solar neutrino sources to
the predicted  event rates for the chlorine and gallium
experiments. The results presented in
table~\ref{tab:radiochemical} were obtained using the best-fit
oscillation solutions of table~\ref{tab:bestfitsa}.  The values
given in the table were calculated with the two-step treatment of
the SNO data. The calculated chlorine and gallium rates are
essentially the same when the best-fit solutions are used that
were obtained including the full SNO day-night energy spectrum. In
that case, the predicted rates for the LMA solution are increased
by about $1$\% and the predicted values for the LOW and vacuum
solutions are decreased by between $2$\% and $7$\%. The rates
shown in table~\ref{tab:radiochemical} are also very similar to
the rates that were expected~\cite{robust} prior to the recent SNO
measurements.

The most important next generation solar neutrino experiment will
detect the fundamental $p$-$p$ neutrinos, which constitute $91$\%
of the total neutrino flux predicted by the BP00 solar model.
Unfortunately, there are no approved $p$-$p$ solar neutrino
experiments at the present time, although there are a number of
very promising proposals under development. Among the CC
detectors, only LENS ~\cite{lens} is in a  advanced stage of
research development and even for LENS important details regarding
the detector are not yet known. There are a number of potential
$\nu-e$ scattering experiments~\cite{lownuexpe} that could observe
$p$-$p$ neutrinos. Therefore, for completeness we have added the
predictions for a generic $\nu-e$ scattering experiment using
$p$-$p$ neutrinos. Table~\ref{tab:afterpredictions} and
table~\ref{tab:beforepredictions} (see also
table~\ref{tab:afterpredictionsnew}) give the calculated values of
the $p-p$ $\nu-e$ scattering rate for two plausible recoil
electron kinetic energy thresholds, $50$ keV and $100$ keV.

We are grateful to the SNO collaboration for the thrill of
analyzing solar neutrino data that include a neutral current
measurement. We have profited from  several discussions with M.
Chen, F. Duncan, and A. McDonald regarding the analysis methods
used by the SNO collaboration and from a constructive referee's
report. JNB acknowledges support from NSF grant No. PHY0070928.
MCG-G is supported by the European Union Marie-Curie fellowship
HPMF-CT-2000-00516.  This work was also supported by the Spanish
DGICYT under grants PB98-0693  and FPA2001-3031.

\end{document}